\documentclass[final,5p,times,twocolumn,numeric]{elsarticle}
\usepackage{amssymb}
\usepackage{lipsum}
\usepackage{cite}
\usepackage{algorithm}
\usepackage[noend]{algpseudocode}
\usepackage{algorithm}
\usepackage{algpseudocode}
\usepackage{float}
\usepackage{booktabs}
\usepackage{amsmath}
\usepackage[hidelinks]{hyperref}
\usepackage{footmisc}

\algnewcommand{\algitem}[1]{\item[]\hspace*{-1em}\textbf{#1}}
\newcommand{\alglistitem}[1]{\item[]\hspace*{1em}\parbox[t]{\dimexpr\linewidth-1em}{\hangindent=1em\strut\textbullet\hspace{0.5em}#1\strut}}

\begin{document}




\title{\textbf{Towards Parsimonious Generative Modeling of RNA Families}}


\author[first,second]{Francesco Calvanese}

\author[second]{Camille N. Lambert}

\author[second]{Philippe Nghe}

\author[third,fourth]{Francesco Zamponi}

\author[first]{Martin Weigt}

\affiliation[first]{organization={Sorbonne Universite', CNRS, Institut de BiologieParis-Seine, Laboratoire de Biologie Computationnelle et Quantitative -- LCQB Paris, France}}

\affiliation[second]{organization={Laboratoire de Biophysique et Evolution, UMR CNRS-ESPCI 8231 Chimie Biologie Innovation, PSL University, Paris, France} }

\affiliation[third]{organization={Dipartimento di Fisica, Sapienza Università di Roma, Rome, Italy}}

\affiliation[fourth]{organization={Laboratoire de Physique de l’Ecole Normale Superieure, ENS, Universite' PSL, CNRS, Sorbonne Universite', Universite' de Paris, Paris, France}}

\begin{abstract}
\textbf{Generative probabilistic models emerge as a new paradigm in data-driven, evolution-informed design of biomolecular sequences. This paper introduces a novel approach, called Edge Activation Direct Coupling Analysis (eaDCA), tailored to the characteristics of RNA sequences, with a strong emphasis on simplicity, efficiency, and interpretability. eaDCA explicitly constructs sparse coevolutionary models for RNA families, achieving performance levels comparable to more complex methods while utilizing a significantly lower number of parameters. Our approach demonstrates efficiency in generating artificial RNA sequences that closely resemble their natural counterparts in both statistical analyses and SHAPE-MaP experiments, and in predicting the effect of mutations. Notably, eaDCA provides a unique feature: estimating the number of potential functional sequences within a given RNA family. For example, in the case of cyclic di-AMP riboswitches (RF00379), our analysis suggests the existence of approximately $\mathbf{10^{39}}$ functional nucleotide sequences. While huge compared to the known $< \mathbf{4,000}$ natural sequences, this number represents only a tiny fraction of the vast pool of nearly $\mathbf{10^{82}}$ possible nucleotide sequences of the same length (136 nucleotides). These results underscore the promise of sparse and interpretable generative models, such as eaDCA, in enhancing our understanding of the expansive RNA sequence space.}
\end{abstract}


\maketitle




\section{Introduction}

RNA molecules play a critical role in many biological processes, including gene expression and regulation. They carry a multitude of functions, such as encoding and transferring genetic information, regulating gene expression, and catalyzing chemical reactions \citep{RNAgeneexpression,RNAroleTranscription,Rybozimes} .

\begin{figure}[htb]
\begin{center}
\vspace{6mm}
\hspace{-7mm}
\includegraphics[width=0.80\columnwidth]{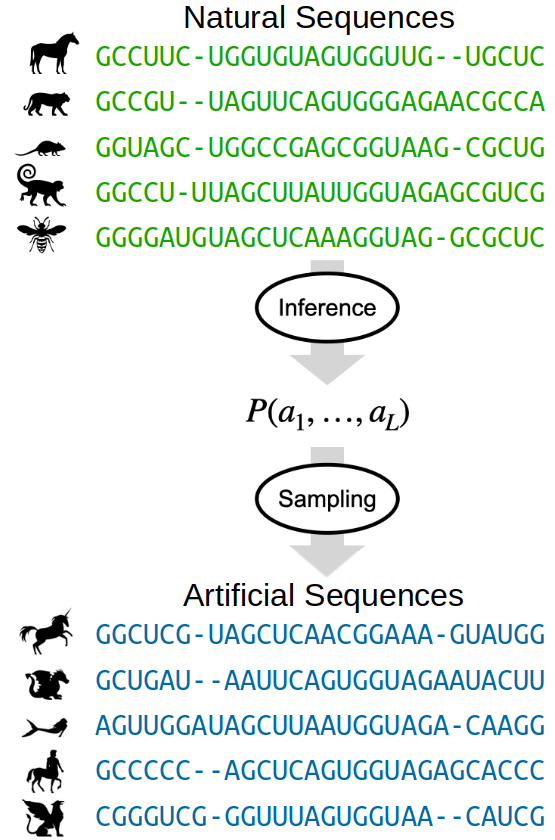}
\end{center}
\caption{Probabilistic generative models extract a probability distribution for the RNA family from natural data, which can then be used to generate artificial sequences. Each of these artificially generated sequences is consistent with the statistics of the RNA family, yet they cannot be attributed to any natural variant, thereby introducing an element of novelty.}
\vspace{0mm}
\label{gen_mod}
\end{figure}

Functional RNA molecules are expected to be extremely rare in the exponentially vast nucleotide-sequence space, and current databases contain only a tiny fraction of the overall possible, functionally viable sequence diversity.
However, it is worth noting that almost identical biological functions can be carried out by different RNA exhibiting substantial sequence variability. Databases like Rfam \citep{Rfam} gather these in diverse yet functionally consistent families of homologous RNA sequences. In computational sequence biology, a significant challenge lies in harnessing the relatively limited pool of existing RNA sequences within a family, often comprising just a few hundred or thousand examples. The objective is to decipher the sequence patterns that underpin the three-dimensional structure and biological functions of these RNA families. This endeavor extends beyond the known sequences, aiming to explore the vast potential space of sequences capable of adopting similar structures and functions. Such analyses provide valuable insights into the complex organization of sequence space and, ultimately, unravel the intricate sequence-to-function relationship. This quest has gained paramount significance, especially in the era of high-throughput sequencing, solidifying its status as one of biology's central and most challenging questions.
Generative probabilistic models offer a powerful approach to tackling these challenges by extrapolating beyond the limited pool of known RNA molecules and generating previously unseen functional sequences. When applied to RNA families, these models build a probability distribution, denoted as $P(a_1,\dots, a_L)$ \citep{eleonora1,weinreb20163d,Pucci,cuturello2020assessing,remisimona}. This distribution encapsulates the variability found in the known sequences within the family while encompassing all possible sequences of length $L$ (for a more precise definition, see {\em Material and Methods}). To provide an intuitive analogy, think of this probability distribution as defining a ``landscape'' across sequence space. Through maximum-likelihood learning, generative models assign high probabilities to sequences considered functional, akin to the peaks in this landscape. Conversely, non-functional sequences receive lower probabilities.
This probability distribution also enables the prediction of mutational effects \citep{fitnessprediction2,Fitnessprediction} since mutations can alter the sequence probabilities relative to the wildtype. Additionally, these models allow for generating novel synthetic sequences \citep{russetal,remisimona} through a sampling process (as illustrated in Fig.~\ref{gen_mod}). A well-constructed model $P$ should possess the ability to generate nucleotide sequences that are diverse but statistically indistinguishable from the known sequences in the family.
Constructing generative models is an exceptionally complex undertaking due to the sheer volume of probabilities they must assign, all while learning from a relatively small pool of existing molecules. As an example, consider an RNA molecule with an aligned length of $L=150$ residues, i.e.~the sequence may contain both nucleotides and gaps. The model must learn approximately $5^L\simeq 10^{105}$ distinct values, even though typical RNA families may consist of only $10^2 - 10^4$ known sequences.
The lack of abundant RNA data makes it hard for complex models like deep architectures to work well, as seen in other tasks like RNA structure prediction \citep{NoAlfaFold}.
This suggests that simpler, less complicated models may be more suited to tackle RNA. 
One of the prominent generative models employed in biomolecular research is the Boltzmann Machine Direct Coupling Analysis (bmDCA) \citep{figliuzzi2018pairwise,muntoni2021adabmdca}. The core idea behind this model lies in the notion that RNA residues of significant functional importance experience evolutionary pressures that deter deleterious mutations. Consequently, these residues tend to remain conserved across the Multiple Sequence Alignment (MSA) collecting homologous sequences. Conversely, pairs of nucleotides that exhibit co-evolutionary patterns over time display correlated mutations. To capture both types of constraints, bmDCA adjusts its probability distribution to mirror the one-site and two-site frequencies observed in the MSA, which serve as proxies for conservation and co-evolution, respectively, cf.~reviews in \citep{de2013emerging,cocco2018inverse}.
In this context, one-site frequencies, denoted as $f_i(a)$, describe how often a nucleotide $a \in { A, U, C, G, - }$ (with '$-$' representing alignment gaps) appears at a specific site $i \in {1,\dots,L}$ within the MSA. Meanwhile, two-site frequencies, denoted as $f_{ij}(a,b)$, provide information about the joint occurrence of nucleotide pairs $(a,b)$ at positions $(i,j)$ within the same sequence. The probability distribution used in bmDCA takes the form of a fully connected Potts/Markov Random Fields model, which captures the interplay of these frequencies,
\begin{equation}\label{P_bmDCA}   
P(a_1,\dots,a_L) = \frac{1}{Z}\exp\left\{{\sum_{i=1}^{L}}h_i(a_i) + \sum_{i<j}J_{ij}(a_i,a_j)\right\} \ ,
\end{equation} 
with $Z$ being the partition function that guarantees normalization. The $h_i(a)$ ($a \in \{A, U, C, G -\}$) are the local `fields' used to fit the one-site statistics. The $J_{ij}(a,b)$ matrices (with $(a,b) \in \{A, U, C, G -\}^2$) are $5 \times 5$ interaction `couplings' used to fit the two-site statistics. 
Although DCA has proven itself as a valuable instrument in investigating proteins, exhibiting achievements in tasks like generating functional sequences \citep{russetal}, forecasting the effect of mutations \citep{fitnessprediction2,Fitnessprediction}, deciphering protein evolution \citep{de2020epistatic,Matteo}, and identifying structural interactions \citep{DCAcontacts,marks2011protein}, its application to RNA remains relatively unexplored \citep{eleonora1,weinreb20163d,Pucci,cuturello2020assessing}. Furthermore, the limited availability of RNA data, compared to the wealth of data for proteins, makes the use of intricate models like large language models \citep{leanguagemodel} impractical. Consequently, employing simpler models for RNA is not only suitable but also presents the benefits of enhanced interpretability, reduced computational burden, and local trainability.
Nonetheless, conventional bmDCA generates a fully connected coupling network (as seen in Eq.~\ref{P_bmDCA}): it models coevolution between all conceivable pairs of residues, even when there is no actual coevolution occurring. As a consequence, this approach can yield a substantial number of noisy couplings $J_{ij}(a,b)$ in the network that lack any statistical support. To mitigate this issue, network sparsification can be applied to trim down the network by eliminating numerous spurious couplings. This process aids in identifying the most informative and functionally significant couplings, rendering the network more accessible for interpretation and analysis. Previous endeavors in this direction have primarily concentrated on sparsifying coupling networks within proteins \citep{sparsification}.
In our work, we introduce a novel approach called Edge Activation Direct Coupling Analysis (eaDCA) specifically tailored for RNA. Unlike previous algorithms, eaDCA takes a unique starting point: an empty coupling network. It then systematically constructs a non-trivial network from scratch, rather than starting with a fully connected network and subsequently simplifying it. This step-by-step process generates a series of models, gradually increasing in complexity until they achieve a statistical performance comparable to that of bmDCA.
Our method offers notable advantages. It operates more swiftly than initiating with a fully connected model, resulting in generative Potts models that demand significantly fewer parameters than standard bmDCA. Furthermore, at each stage of our approach, we employ analytical likelihood maximization. This feature allows us to easily track normalized sequence probabilities and estimate entropies throughout the network-building process. This invaluable information enhances our ability to interpret and analyze the vast space of RNA sequences.
The organization of the manuscript is as follows. In `Materials and Methods', we present the foundational principles and functionality of the model, describe the data used in the model training and analysis, and provide specific information about the SHAPE-MaP experiments conducted to examine artificial molecules.
In `Results and Discussion', we evaluate the statistical properties of the artificial sequences generated by eaDCA, interpret the parameters of the sparse architectures, and examine the model's predictions regarding mutational effects on tRNA. Additionally, using eaDCA to access normalized sequence probabilities and model entropies, we conduct an analysis on how different constraints, such as compatibility with secondary structures or family conservation and coevolution statistics, affect the size of the viable RNA sequence space. Lastly, we assess the SHAPE-MaP experimental results, characterizing the structure of artificially generated tRNA molecules.

\section{Materials And Methods}
In this section we discuss the data and methodological basis of our work: the data used for training and evaluating our models, the new algorithm proposed here, and the experimental protocol to test artificial sequences generated by our approach.
\subsection{\textbf{Data}}
\subsubsection{RNA families --}
All generative models discussed here are trained for individual RNA families, i.e.~homologous but diverged sequences of largely conserved structure and function \citep{Rfam}. Each family is represented by a Multiple Sequence Alignment (MSA) $\mathcal{D} = (a_i^r \,|\, i = 1,\dots,L;\, r = 1,\dots,M)$, with $L$ indicating the aligned sequence length, and $M$ the number of distinct sequences. The entries $a_i^r$ are either one of the four nucleotides $\{A,C,G,U\}$, or the alignment gap ``--'' reflecting insertions and deletions in the original unaligned sequences.
Following standards in the literature, phylogenetic effects are partially compensated by reweighting each sequence by a factor $\omega_r$ \citep{cocco2018inverse}, which equals the inverse number of all sequences having more than 80\% sequence identity to sequence $r$, and which is used when estimating the empirical single-site nucleotide frequencies $f_i(a)$ and pair frequencies $f_{ij}(a,b)$ from the data $\cal D$, cf. the supplementary information (SI) for details. The sum of weights $M_{\rm eff}=\sum_r \omega_r$ defines the effective sequence number as a more accurate reflection of the diversity of the dataset.

eaDCA is tested on $25$ RNA families of known tertiary structure with $L$ ranging from $50$ to $350$ and $M$ from $30$ to $50000$. These families are extracted from the CoCoNet benchmark dataset \citep{CoCoNet} by limiting ourselves to datasets with high $M_{\rm eff}$ and sequence length $L<350$. The MSA were updated using a more recent Rfam release (May 2022), and matched to exemplary PDB structures. A comprehensive list of family names, characteristics, and used PDBs is given in Table~\ref{table} of the SI. 

The main text concentrates on two families: the tRNA family (RF00005) was selected due to the existence of mutational datasets, and our own experiments were performed on this family. Due to its unusually large size, the MSA was randomly downsampled to $M=30,000$ sequences. The cyclic di-AMP riboswitches (RF00379) were chosen due to their interesting and non-trivial statistical properties. The robustness of all results is illustrated in the SI, where the other 23 families are exhaustively analyzed.

\subsubsection{Mutational Fitness Dataset --}
To evaluate our ability to predict mutational effects in RNA molecules, we utilized the data published in \citep{tRNAfitness}. 
This dataset provides {\em in vivo} fitness measurements for $23,283$ variants of the yeast tRNA$_{\text{Arg}}^{\text{CCU}}$ at temperatures of $23^{\circ}C$, $30^{\circ}C$, and $37^{\circ}C$, with up to 10 mutations compared to wildtype. These mutations may result in non-functional sequence variants, in difference to the natural sequences in the RNA families. We focus on the results at $37^{\circ}C$ because, at higher temperature, the tRNA$_{\text{Arg}}^{\text{CCU}}$ becomes increasingly important for the survival of the organism. The details of the datasets and the results for $23^{\circ}C$ and $30^{\circ}C$ are provided in the SI. Fitness values are organized such that $0.5$ represents a mutant yeast strain incapable of reproduction, while $1.0$ is the wildtype fitness.

\subsubsection{SHAPE Reference Dataset --} 
In order to empirically validate our generative models, we conducted Selective 2’-Hydroxyl Acylation analyzed by Primer Extension with Mutational Profiling (SHAPE-MaP) experiments on artificially generated tRNA molecules, cf.~below.
To ensure the robustness of our analysis and to facilitate a meaningful comparison, we utilized an external published dataset comprising SHAPE reactivity profiles for $20$ RNA sequences with known secondary structure. This dataset, which we will refer to as the `SHAPE Reference Dataset', was obtained from \citep{SHAPEreference} . 

\subsection{\textbf{Edge Activation Direct Coupling Analysis (eaDCA)}}

\subsubsection{Algorithm Principle --}
The proposed algorithm belongs to the family of DCA algorithms, i.e.~it learns a Potts model in the form of Eq.~\ref{P_bmDCA} from an MSA ${\cal D}$. However, instead of introducing couplings $J_{ij}(a,b)$ for all pairs of nucleotide positions $0\leq i<j\leq L$, we aim at a parsimonious model and activate couplings only between those pairs, which are really coevolving and thus essential for the accurate statistical description of the sequence family. All other pairs, which do not have clear signatures of direct coevolution, shall not be included into the set of coevolutionary couplings, to avoid noise overfitting \citep{sparsification} .

\begin{figure}[t] 
\begin{center} 
\includegraphics[width=1.\columnwidth]{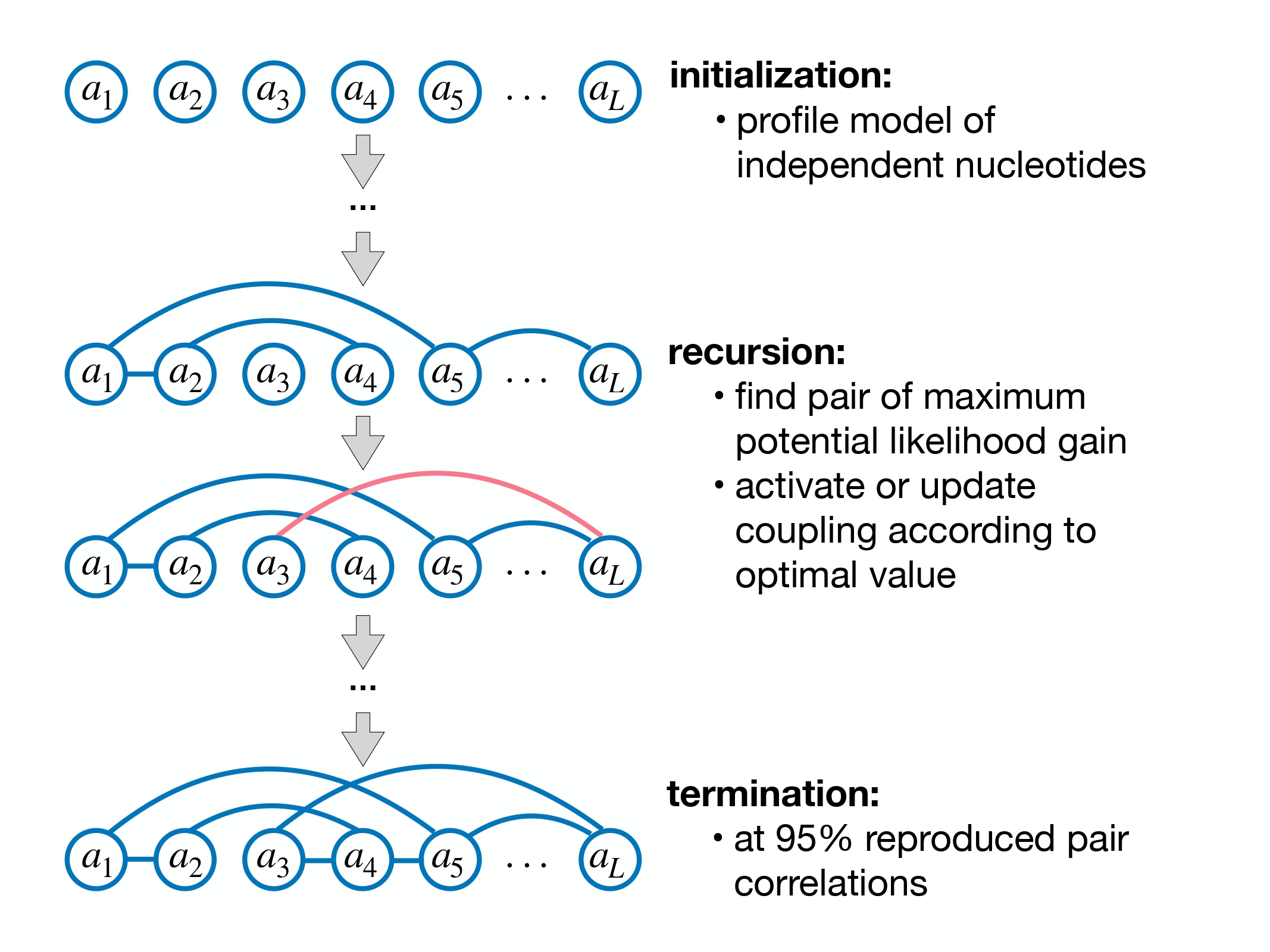}
\end{center}
\caption{Schematic representation of the recursive eaDCA algorithm.}
\vspace{0mm}
\label{proced}
\end{figure}

Since the empirical pair frequencies $f_{ij}(a,b)$ are shaped both by direct coupling and indirect correlation, the set of coupled pairs, ${\cal E}=\{(ij) \; | \; J_{ij} \text{ is non-zero}\}$, cannot be fixed in a single step, but has to be constructed recursively, as is shown schematically in Fig.~\ref{proced} and detailed below: starting from a profile model of independent nucleotides, ${\cal E}_0= \emptyset$, we construct a series of edge sets ${\cal E}_t$, by activating or updating edges one by one. In this setting, ``activating'' an edge signifies to introduce a non-zero coupling for a previously uncoupled pair $(ij)$, while ``updating'' indicates a change of the coupling value on an already activated edge. As a consequence, at any algorithmic step $t$, the model can be written as 
\begin{eqnarray}\label{Pt}   
P_t(a_1,\dots,a_L) &=& \frac{1}{Z_t}\exp\left\{ -E_t(a_1,\dots,a_L)\right\} \nonumber\\
E_t(a_1,\dots,a_L) &=& -
\sum_{i=1}^L h_i(a_i) - \sum_{(ij)\in{\cal E}_t}J_{ij}(a_i,a_j) \ ,
\end{eqnarray}
with $E_t$ being called ``statistical energy''. The log-likelihood of the model given the reweighted data $\cal D$ reads 
\begin{equation} \label{llt}
\mathcal{L}_t = \sum_{r=1}^M \omega_r \log P_t(a_1^r,\dots,a_L^r) \ .
\end{equation}

\subsubsection{Initialization --}
As already mentioned, the model is initialized without couplings, ${\cal E}_0= \emptyset$, and reads 
\begin{equation}\label{P0}   
P_0(a_1,\dots,a_L) = \frac{1}{Z_0}\exp\left\{{\sum_{i=1}^{L}}h_i(a_i) \right\} \ .
\end{equation} 
The log-likelihood ${\cal L}_0$ is easily maximized by setting
\begin{equation}\label{h}   
h_i(a) = \log f_i(a)
\end{equation} 
for all $i=1,...,L$ and $a\in\{A,C,G,U,-\}$, i.e.~the model reproduces the empirical single-residue statistics. The resulting partition function is $Z_0=1$. This simple model is known under the name of profile model (or independent-site model) and widely used in bioinformatic sequence analysis.

\subsubsection{Recursion --}
The algorithmic step from $t$ to $t+1$ is characterized by a modification of a single $5\times 5$ coupling matrix $J_{kl}$ on a single position pair $(kl)$,

\begin{eqnarray}
\hspace{0.3cm}
            E_{t+1}(a_1,...,a_L) &=& E_{t}(a_1,...,a_L) - \Delta J_{kl}^*(a_k,a_l) \ ,
    \nonumber\\
    {\cal E}_{t+1} &=& {\cal E}_t \cup \{(kl)\} \ .
\end{eqnarray}

If $(kl)$ was not yet active in ${\cal E}_t$, this corresponds to an edge activation, otherwise to an edge update.

The edge $(kl)$ and the coupling change $\Delta J_{kl}^*(a_k,a_l) $ are chosen to maximize the log-likelihood ${\cal L}_{t+1}$. As is proven in the SI, this is realized by choosing the pair 
\begin{equation} \label{DKL}
    (kl) = \operatorname*{argmax}_{1\leq m < n \leq L} \,D_{KL}\left(f_{mn} \,||  \, P^t_{mn}\right) \ ,
\end{equation}
i.e.~the currently least accurate position pair, in which the current model's marginal two-residue distribution $P^t_{mn}$ deviates most from the empirical target distribution $f_{mn}$. $D_{KL}$ denotes the standard Kullback-Leibler divergence,
\begin{equation}
    D_{KL}\left(f \,||  \, P\right) = \sum_{a,b} \, f(a,b)\, \log \frac{f(a,b)}{P(a,b)} \ ,
\end{equation}
for any pair of probability distributions $f$ and $P$. Note that the selection goes over all position pairs $m,n$, independently on their activation status in ${\cal E}_t$. Note also that the exact determination of the marginal distributions $P^t_{mn}(a,b)$ is infeasible since it would require to sum over all $5^L$ possible sequences of aligned length $L$. We therefore use Markov chain Monte Carlo (MCMC) sampling; the exact procedure based on persistent contrastive divergence is detailed in the SI.

The optimal coupling change is also derived in the SI, it equals the log-ratio of the empirical and the current model probabilities on edge $(kl)$,
\begin{equation} \label{update}
    \Delta J^*_{kl}(a,b) = \log \frac{f_{kl}(a,b)}{P^t_{kl}(a,b)}\ .
\end{equation}
To avoid excessively high values for rare amino-acid combinations, this term is regularized using pseudocounts for both the empirical frequencies and the model probabilities, cf.~the SI for details.

\subsubsection{Termination --}
As the  process continues, the resulting models become increasingly accurate and complex. It can be observed from Eq.~(\ref{update}) that a fixed point is attained when all the two-point probabilities are equal to their respective empirical frequencies. This corresponds exactly to the fixed-point condition imposed in bmDCA. Because this condition is impossible to achieve in practice due to MCMC sampling noise, we set an {\em ad hoc} stopping criterion by looking at how well the empirical two-site covariances 
\begin{equation} \label{cij}
    c_{ij}(a,b)=f_{ij}(a,b)-f_i(a)f_j(b)
\end{equation}
are reproduced by the connected correlations in the model,
\begin{equation} \label{cijt}
    c_{ij}^t(a,b)=P_{ij}^t(a,b)-P_i^t(a)P_j^t(b)\ .
\end{equation}
The algorithm terminates at step $t_f$ when the Pearson correlation $\rho$ between these two quantities, evaluated over all positions $i,j$ (including those not in ${\cal E}_{t_f}$) and all nucleotides $a,b$ (including gaps), reaches 0.95. This value is commonly reached in bmDCA as well. The reason for computing the score based on the $c_{ij}(a,b)$ instead of the $f_{ij}(a,b)$ is that the former isolates coevolution statistics from the conservation ones.

The entire procedure is summarized as a pseudocode in Alg.~\ref{pseudocode}, and represented graphically in Fig.~\ref{proced}.

\begin{algorithm}[H] 
\caption{(eaDCA)}
\label{pseudocode}
\begin{algorithmic}
\algitem{Initialization:}
    \alglistitem{Profile model: $P_0(a_1, \dots, a_L) = \prod_{i=1}^Lf_i(a_i)$}
    \alglistitem{Iteration counter: $t \gets 0$}
\algitem{Recursion:}
\algitem{While} $\rho (c^t_{ij},c_{ij})< 0.95$ \textbf{do}
    \alglistitem{Estimate two-point probabilities $P^t_{ij}(a,b)$ for all $i, j$ and all $a,b$ via MCMC sampling}
    \alglistitem{Identify the worst represented edge $(kl)$ according to Eq.~(\ref{DKL})}
    \alglistitem{Update the interaction on the identified edge using Eq.~(\ref{update}) to get the new model $P_{t+1}(a_1, \dots, a_L)$}
    \alglistitem{Add the identified edge $(kl)$ to ${\cal E}_t$ to get ${\cal E}_{t+1}$}
    \alglistitem{Increment iteration counter: $t \gets t + 1$}
\algitem{Termination} at $t_f = t$ :
\alglistitem{Output $P_{t_f}(a_1,\dots,a_L)$ with $95\%$ reproduced pair correlations}
\end{algorithmic}
\end{algorithm}

\subsubsection{Normalized Sequence Probabilities and Model Entropy~--} Probabilistic generative models typically do not provide normalized sequence probabilities but only relative sequence weights. 
This limitation arises because obtaining normalized probabilities would necessitate summing over the entire $5^L$
sequence space to get the partition function $Z$ given by Eq.~(\ref{P_bmDCA}), 
\begin{equation}   
Z=\sum_{a_1,\dots,a_L}\exp\left\{{\sum_{i=1}^{L}}h_i(a_i) + \sum_{(ij) \in {\cal E}}J_{ij}(a_i,a_j)\right\} \ ,
\end{equation}
which is infeasible for any biologically relevant value of $L$. Relative weights are sufficient for MCMC sampling of artificial sequences, but they are meaningful just within the context of a specific model and cannot be compared across distinct models.

The advantage of eaDCA is that the recursion preserves model's partition function $Z$, as is shown in SI. Since $P_0$ is trivially normalized, we have
\begin{equation}
Z_0=1 \ \ \text{ and } \ \ \ Z_{t+1}=Z_t \ ,
\end{equation}
i.e.~the models remain trivially normalized under recursion:
\begin{equation}
    P_t(a_1,\dots,a_L)= \exp\left\{ -E_t(a_1,\dots,a_L) \right\} \ .
\end{equation}
A nice consequence of this propriety is that we have easy access to the model's entropy $S_t$
\begin{eqnarray}\label{entr}
\hspace{1.4cm}
S_t &=& -\langle \log P_t(a_1,\dots,a_L) \rangle_{P_t}\\
    &=&  \langle E_t(a_1,\dots,a_L) \rangle_{P_t}
\end{eqnarray}
via the average statistical energy, which can be accurately estimated from an MCMC sample. From the entropy $S_t$ we can deduce the size of the viable sequence space, 
\begin{equation} \label{Omega}
    \Omega_t = \exp \left\{S_t\right\} \ ,
\end{equation}
which can be thought of as the effective number of different sequences that we can sample from $P_t(a_1,\dots,a_L)$.

In practice, because we depend on stochastic MCMC techniques for estimating the two-site probabilities $P_{ij}(a,b)$ in eaDCA iterations, the $Z$ is only approximately conserved. However, it is straightforward to accurately monitor and account for these errors, see the SI for details.

\subsection{\textbf{SHAPE-MaP probing of artificial tRNA molecules}}

To conduct an empirical evaluation of our eaDCA-derived model, we performed a SHAPE-MaP analysis on a set of $76$ artificially generated tRNA molecules (RF00005 family).  
Here we summarize the experimental protocols, full details are provided in the SI.

\subsubsection{RNA production --}
We designed a total of 76 tRNAs. Each RNA was synthesized with the T7 promoter positioned at its 5' end and the last 16 nucleotides were kept constant for all constructs matching those of the yeast tRNA(asp). The DNA templates (gBlock or oligoPools from Integrated DNA Technologies) were amplified by PCR using the Phusion Hot Start Flex polymerase (New England Biolab). After purification, the DNAs were transcribed via in vitro transcription using the HiScribe T7 High Yield RNA Synthesis Kit (NEB). The resulting RNAs were purified by denaturing gel electrophoresis.

\subsubsection{RNA modification --}
The SHAPE reactivity is not only a reflection of RNA structure 
but also depends on experimental conditions, necessitating careful consideration 
in comparative analysis of SHAPE-MaP reactivity profiles \citep{SHAPEint} . 
Consequently, we chose to probe our artificial tRNA with the same folding buffer 
(50 mM HEPES pH 8.0, 200 mM potassium acetate pH 8.0, and 3 mM MgCl2)
than the yeast tRNA(asp) Reference SHAPE Dataset \citep{SHAPEreference} .
For RNA modification, three conditions were performed: positive (with the probe), negative (only the probe solvent) and denaturing (denatured RNA with the probe). For the positive and negative conditions, the RNAs were allowed to refold and the modifying agent (1M7 in DMSO for positive) or the solvent (neat DMSO for negative) were quickly mixed to the RNAs and incubated 5 min at 37°C. For the denaturing condition, the RNAs were first denatured by addition of formamide followed by a heat treatment and the RNAs were modified similarly (1M7 probe). After incubation, all modified RNAs were purified via ethanol precipitation and quantified by the Qubit RNA High Sensitivity assay kit (ThermoFisher).

\subsubsection{Library preparation --}
The modified RNAs were pooled in equimolar proportion based on their conditions (positive, negative, denaturing) and reverse-transcribed using the SuperScript II reverse transcriptase (ThermoFisher) with a buffer allowing the misincorporation of nucleotides at the chemically modified positions. We also used a Template Switching Oligo (TSO) in order to incorporate the Rd1 Illumina adapter during the reverse-transcription, and brought the Rd2 Illumina adapter by the reverse-transcription primer. After cDNA purification, PCR enrichment was conducted to amplify the DNA libraries and incorporate the P5/P7 Illumina adaptors. The samples were purified by AMPure XP beads (Beckman Coulter), quantified by quantitative PCR (KAPA Library Quantification Kit, Roche), and sequenced on a MiSeq-V3 flow cell (Illumina) at the NGS platform of Institut Curie (Paris, France).

\subsubsection{SHAPE Reactivity Mapping --}
We employed the ShapeMapper2 \citep{ShapeMapper} software to process the sequencing data, obtaining SHAPE reactivity values for each artificial tRNA molecule, which partition sites into the reactivity classes `low', `medium' and `high' \citep{RNAstructure,SHAPEint}. ShapeMapper2 was run with default settings, except for the depth-per-site quality threshold that was lowered from $5000$ to $3000$. This allowed us to gather reactivity data covering more than $50\%$ or the residues for at least $30$ of the molecules under investigation.

\section{Results And Discussion}

\subsection{\textbf{eaDCA models reproduce the natural sequence statistics}}

The initial evaluation of the performance of any generative model involves assessing its ability to accurately replicate the statistical properties of natural sequences. For this, we conduct an analysis across $25$ RNA families, comparing the statistical properties of their natural sequences with those of independently and identically distributed (i.i.d.) samples, generated from both the eaDCA model and a simpler, secondary-structure based covariance model (CM) \citep{CovMod}. In the latter, only nucleotide pairs involved in secondary structure (S2D) are connected by couplings. The corresponding CM can be written as a Potts model, with all maximum-likelihood parameters given exactly by the empirical one- and two-nucleotide statistics,
\begin{equation} \label{CM}
\begin{gathered}
P_{CM}(a_1,\dots,a_L) = \exp\left\{ \sum_{i=1}^{L} h_i(a_i) + \sum_{(ij)\in S2D} J_{ij}(a_i,a_j)\right\} \\
J_{ij}(a_i,a_j)=\log \frac{f_{ij}(a_i,a_j)}{f_i(a_i)f_j(a_j)},\quad \quad h_i(a_i)=\log f_i(a_i) \ .
\end{gathered}
\end{equation}
We use the CM as a performance benchmark over the classical profile model because the information about RNA secondary structure is readily available, and because the base-pair complementarity in RNA secondary structure causes a strong pairwise coevolution. Also, CM are at the basis of Rfam MSA, since they are used in RNA homology detection and sequence alignment by Infernal \citep{Infernal}.

\begin{figure}[h!] 
\begin{center}
\hspace*{-0.4cm}
\includegraphics[width=1\columnwidth]{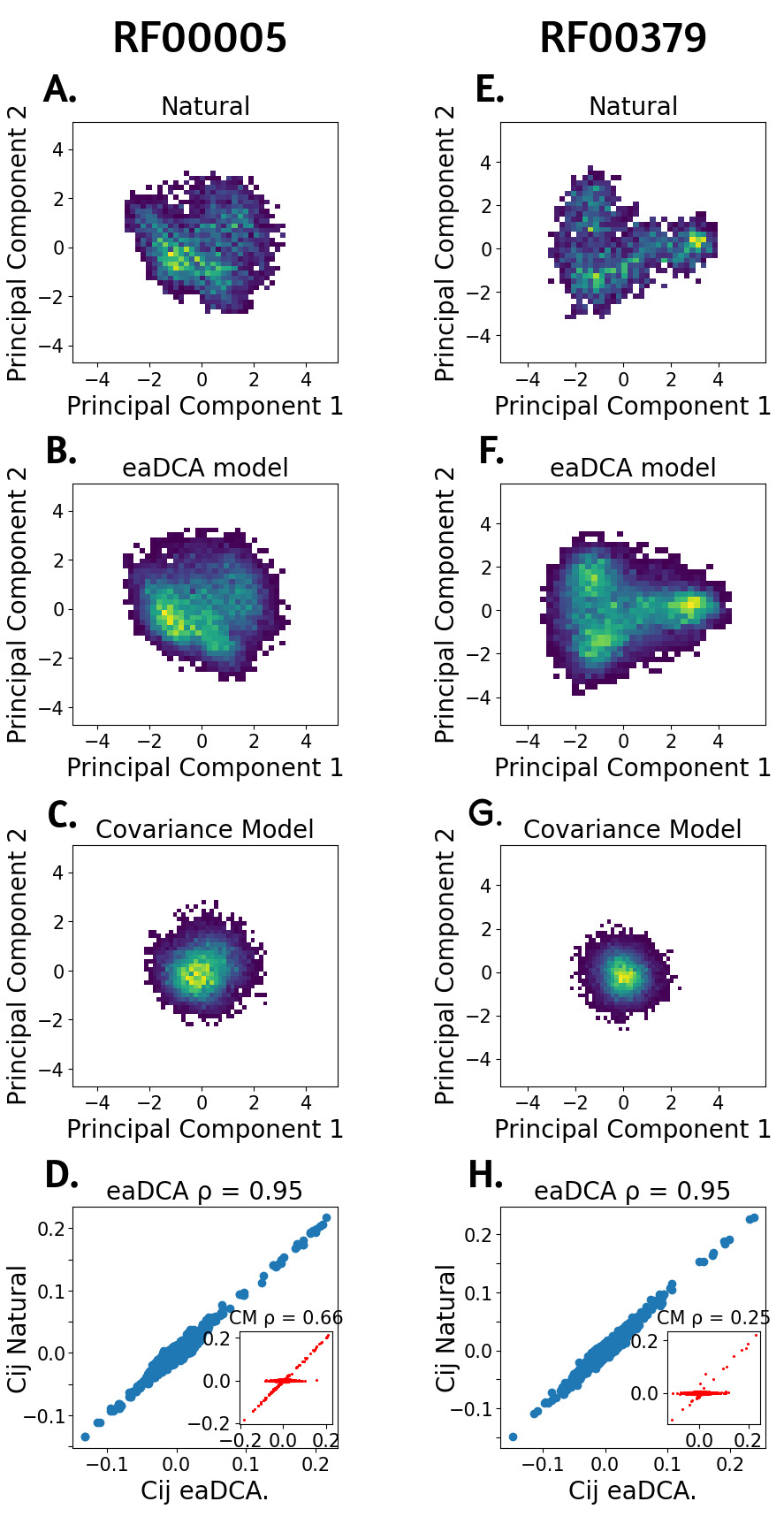}
\end{center}
\caption{\textbf{A.} RF0005: Principal component analysis of natural sequences ($M=28770$). \textbf{B.} RF005: eaDCA generated sequences mapped to the first two principal components of the natural sequences ($M=12000$). \textbf{C.} RF005: CM generated sequences mapped to the first two principal components of the natural sequences ($N=12000$).  \textbf{D.} RF00005: scatter plot of the connected two-site correlations  of the natural sequences vs. eaDCA generated sequences (blue) or CM generated sequences (red - insert) ($N=12000$). \textbf{E.} RF00379: Principal component analysis of natural sequences ($N=3808$). \textbf{F.} RF00379: eaDCA generated sequences mapped to the first two principal components of the natural sequences ($N=12000$). \textbf{G.} RF00379: CM generated sequences mapped to the first two principal components of the natural sequences ($N=12000$).  \textbf{H.}  RF00379: scatter plot of the connected two-site correlations  of the natural sequences vs. eaDCA generated sequences (blue) or CM generated sequences (red - insert) ($N=12000$).}
\label{PCA}
\end{figure}

Fig.~\ref{PCA} displays statistical analyses for the RF00005 and RF00379 families. Additional results for 23 other RNA families can be found in the SI. Figs.~\ref{PCA}D and \ref{PCA}H  display the comparison between the connected two-point correlations of the natural data with estimates from a sample of an eaDCA obtained model and the CM. The results indicate a strong correlation of eaDCA with the natural data for all residue pairs, including those not connected by activated edges, while the CM reproduces the pair correlations only on the secondary structure, and totally fails on all other pairs of positions.

A second test of the eaDCA model's generative properties is demonstrated in Figs.~\ref{PCA}A-C, and \ref{PCA}E-G, which present the natural, eaDCA, and CM-generated sequences projected onto the first two principal components (PCs) of the natural MSA \citep{russetal,autoregressive}. The sequences sampled from the eaDCA model effectively reproduce the visible clustered structure of the natural sequences, while CM are unable to do so, with projections on the PCs being concentrated around the origin. 

The observations in Fig.~\ref{PCA} indicate the inability of CM to serve as accurate generative models, while sequences sampled from eaDCA are coherent with the natural data on the tested observables.

From Table~\ref{table}, we conclude that eaDCA delivers  generative models able to reproduce the natural RNA statistics with only a fraction of the number of parameters of a standard bmDCA implementation (parameter reduction of 84.85\% for RF00005 and  87.83\% for RF00379). A complete table for all the $25$ families is provided in the SI and confirms this observation across families.

\begin{table*}
\centering
\caption{eaDCA results for RF00005 and RF00379 at termination $t=t_f$.}
\label{table}
\begin{tabular}{|c|c|c|c|c|c|c|c|c|c|}
 \toprule
 Name  & $L$ &  $M$ & $M_{eff}$ & CM $c_{ij}$ corr & eaDCA $c_{ij}$ corr  & Parameter Reduction (PR$\%$) & $S$ & $\Omega$ \\ 
\midrule
RF00005 & $71$  & $28770$ & $2267$ & $0.66$ & $0.95$  & $84.85\%$ & $51.34$ & $1.98\times 10^{22}$  \\
RF00379 & $136$ & $3808$  & $1428$  & $0.25$ & $0.95$  & $87.83\%$ & $89.56$ & $1.05\times 10^{39}$  \\
\hline
\end{tabular}
\end{table*}

\subsection{\textbf{Parameter interpretation}}

\begin{figure}[t]
\begin{center}
\hspace*{-0.8cm}
\includegraphics[width=1.1\columnwidth]{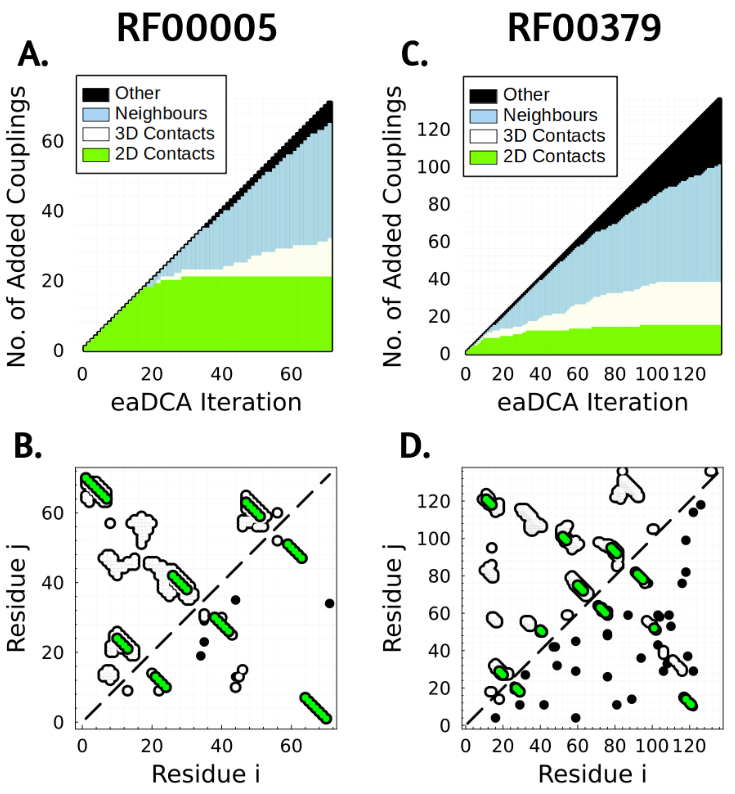}
\end{center}
\caption{\textbf{A.} RF0005: first $L$ activated edges colored according to their classification. \textbf{B.} RF0005: contact map (upper-left) and activated edges (lower-right). Secondary-structure contacts are evidenced in green, non-contacting activated edges in black. \textbf{C.} RF00379: first $L$ activated edges colored according to their classification. \textbf{D.} RF00379: contact map (upper-left) and activated edges (lower-right). Secondary-structure contacts are evidenced in green, non-contacting activated edges in black.}
\vspace{0mm}
\label{edges}
\end{figure}

A key benefit of employing a parsimonious generative model is the potential for obtaining a more insightful interpretation of its parameters. In the context of RNA, the eaDCA method is producing good generative models with a small percentage of the parameters of fully connected models (bmDCA), which in turn enables easier biological interpretation. Since the edge activation procedure starts from the profile model, all single-site frequencies $f_i(a)$ are accurately reproduced from the beginning. Due to its iterative nature, eaDCA produces additionally an ordered list of edges carrying non-zero couplings. These added edges can be used to explain the connected two-point statistics to high accuracy, and they are thus carrying the full information about residue coevolution in the MSA of the RNA family under consideration. 

For this study, we classified the first $L$ added edges into four categories: `secondary structure base pairs' ($S2D$), `tertiary structure contacts' (if the distance between the involved residues is less than $8 \, \text{\AA}$), `neighbors' (if the pair is less than four positions apart along the primary sequence), and `other' (not fitting into any of the prior categories). We present here the analyses for two RNA families (RF0005 and RF00379) but the results of all $25$ families can be found in the SI.

In Fig.~\ref{edges}, the analysis revealed a relationship between contacts and added edges, with almost all $S2D$ pairs  being systematically taken in the early iterations. This trend is consistent with their strong coevolutionary relationship, and shows that CM models capture many of the strongest, but by far not all such relationships. Tertiary contacts are included much later (and many never activated even at termination of the algorithm); we therefore conclude that they typically induce a much lower coevolutionary signal than secondary-structure contacts. The presence of activated edges between neighboring residues may in part be attributable to phylogenetic relationships, but also to the insertion or deletion of multiple nucleotides, i.e.~to the presence of gap stretches in the MSA. 

A relatively small fraction of activated edges do not offer an interpretation (class "other"), it remains unclear if these edges reflect the limited statistics in the natural MSAs, or coevolution beyond structural contacts. eaDCA considers them important for reproducing the natural sequence statistics. In this context, it is important to note that the complete list of edges generated by eaDCA before meeting the termination condition significantly exceeds the sequence length $L$, consequently leading to a large quantity of `other' entries.

\subsection{\textbf{Prediction of mutational effects}}

Potts models (including profile, CM and DCA models) are energy-based statistical model, cf.~Eq.~(\ref{P_bmDCA}). The maximum-likelihood strategy used in their training assumes that nicely functional sequences have high probability, or equivalently low energy. Conversely, low-probability / high-energy sequences do not obey the evolutionary constraints learned by the model, and are expected to be non-functional. 

This property can be used to predict mutational effects \citep{Fitnessprediction, fitnessprediction2} by comparing the energies of the mutated and the wildtype sequences. In this way, a mutant sequence can be characterized by the energy difference
\begin{equation*}
\Delta E = E(\text{mutant}) - E(\text{wildtype})\ .
\end{equation*}
A positive $\Delta E$ implies a reduction in the model probability for the mutant, suggesting that the mutation is likely to be deleterious. On the contrary, a negative $\Delta E$ signals a potentially beneficial mutation. 

\begin{figure}[h!]
\begin{center}
\includegraphics[width=1\columnwidth]{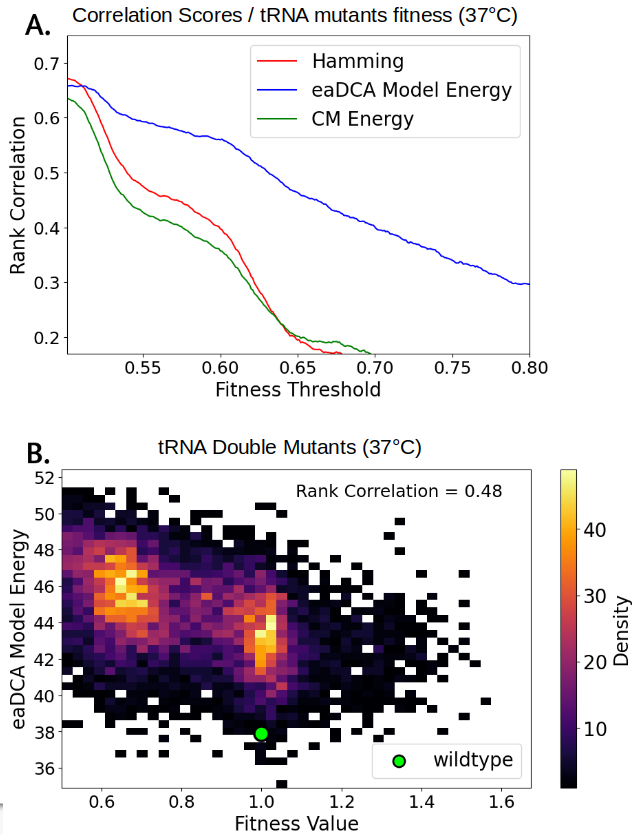}
\end{center}
\caption{\textbf{A.} Correlation of Hamming distance, eaDCA model energy and CM energy with tRNA fitness at different values of minimum fitness threshold $f_\theta$. \textbf{B.} Relation between eaDCA model energy and tRNA fitness for the $8101$ double mutants.  }
\vspace{0mm}
\label{mutations}
\end{figure}

To test the quality of these predictions, we use the tRNA fitness dataset \citep{tRNAfitness} discussed above in {\em Materials and Methods}. We perform the following steps:
\begin{itemize}
    \item For all mutant sequences in this dataset, we determine the energy differences to wildtype using both the eaDCA model, $\Delta E_{eaDCA}$, and the covariance model, $\Delta E_{CM}$, as well as the Hamming distances (i.e.~the number of mutations from wildtype).
    \item We select all mutant sequences having experimental fitness values $f\geq f_\theta$ above an arbitrary fitness threshold $f_\theta$. This threshold is varied in our analyses to focus on diverse strengths of mutational effects.
    \item We calculate the Spearman rank correlation between the three predictors (eaDCA, CM, Hamming) and the fitness values $f$ over the selected mutants, as functions of the fitness threshold $f_\theta$.
\end{itemize}
As is shown in Fig.~\ref{mutations}A, when all mutants are included ($f_\theta = 0.5$), all three predictors show similarly good correlation values between 0.6 and 0.7. This results from the fact that most higher-order mutants, i.e.~those of higher Hamming distance, have very low fitness, while mutants with one or two mutations frequently show more moderate fitness values. However, when increasing the fitness threshold $f_\theta$, i.e.~when including only mutations of more moderate fitness effects, $\Delta E_{eaDCA}$ correlations remain much more robust while the other two rapidly decay with $f_\theta$. This shows that the eaDCA energies are informative over variable ranges of fitness effects.

To corroborate this finding, Fig.~\ref{mutations}B shows a heatmap of the $8101$ two-point mutant sequences (at fixed Hamming distance of 2), comparing $\Delta E_{eaDCA}$ predictions and fitness values $f$. We observe a robust correlation even in this case, where the Hamming distance is constant and thus uncorrelated to the fitness measures.

\

\subsection{\textbf{Size estimation and constraint analysis of RNA sequence space}}

\begin{figure}[h!]
\begin{center}
\hspace*{-1.4cm} 
\includegraphics[width=1.2\columnwidth]{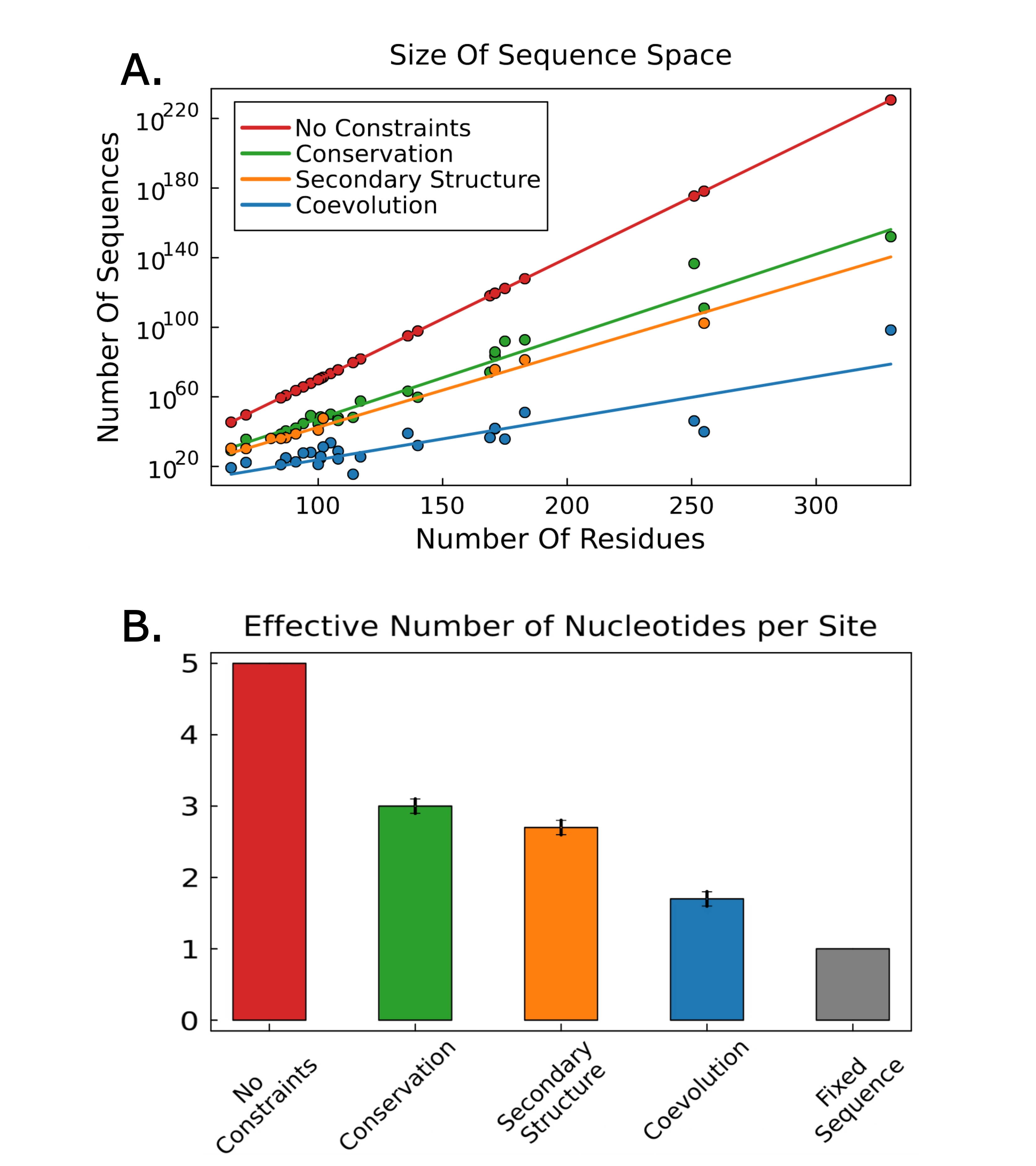}
\end{center}
\caption{\textbf{A.} Relations between RNA family length and the size of the sequence space under coevolution, conservation and secondary structure constraints. Dots are results for the different RNA families studied in this work, and lines are exponential fits. \textbf{B.} Effective number of nucleotides per site $x$ for each constraint. The size of the compatible sequence space is $\Omega = x^ L$.}
\vspace{0mm}
\label{entropy}
\end{figure}
The entire space of sequences of given length is enormous. To illustrate this, the number of all ungapped sequences of length $L=150$ is $4^{150}\simeq 2\times 10^{90}$, if we include gaps like in our MSA, the number even rises to $5^{150}\simeq 7\times10^{104}$, and this exceeds by 10-24 orders of magnitude the estimated number $\sim 10^{80}$ of atoms in the universe. However, the viable sequence space related to a specific RNA family, i.e.~to all sequences taking similar structure and performing similar function, is expected to be much smaller: sequences have to meet constraints imposed by residue conservation and coevolution, and possibly by other evolutionary constraints.

Our models allow for analyzing the impact of the different constraints on the entropy $S$ and the size $\Omega=e^S$ of the sequence space, using the approach discussed in {\em Materials and Methods}. More specifically, the influence of conservation is measured via the entropy $S_0$ of the initial profile model, while the combined influence of conservation and coevolution is measured via the entropy $S_{t_f}$ of the final model at termination \citep{entropy}. These results are corroborated by an independent estimation using a code published in \citep{S2Dsize}, which estimates the size of the sequences space compatible with a given secondary structure, by efficiently sampling the neutral network related to a given RNA secondary structure.

The results are shown in Fig.~\ref{entropy}A for our selected RNA families. All three constraints enforce an exponential relationship between the size of the sequence space $\Omega$ and the sequence length $L$, i.e.~the per-site reduction of the sequence space due to any individual type of constraint is roughly constant across the tested RNA families. Interestingly, conservation and secondary structure constrain the sequence space similarly, while the constraints imposed by both conservation and coevolution are, in line with expectations, the most stringent one. As illustrated in Fig.~\ref{entropy}B, out of the initially $5^L$ possibly gapped sequences of aligned length $L$ about $(2.98 \pm 0.10)^L$ are compatible with the empirical conservation statistics, $(2.66 \pm 0.09)^L$ with the consensus secondary structure of the RNA families, and finally $(1.74 \pm 0.09)^L$ with both conservation and coevolution. To go back to our initial example $L=150$, the final eaDCA sequence space would contain about $10^{36}$ distinct sequences: this number, while remaining enormous as compared to the observed extant sequences found in sequence databases like Rfam, comprises only a tiny fraction of $~10^{-68}$ of the entire sequence space of this length, illustrating the fundamental importance of such constraints in the natural evolution of RNA families.

Note that these numbers also have an interesting interpretation in terms of the effective number of nucleotides, which are, on average, acceptable in a typical position of a functional RNA molecule. Out of the 5 theoretical possibilities (4 nucleotides or an indel), close to three are compatible with familywide  conservation, or 2.66 with the consensus secondary structure. However, both constraints are insufficient for generative modeling as shown before. Our generative modeling indicates a much stronger reduction of the effective number of acceptable nucleotides per site to only 1.74 on average.

\subsection{\textbf{Structural characterization of artificial tRNA molecules by SHAPE-MaP probing}}
\begin{figure}[h!]
\begin{center}
\hspace*{-0.8cm}
\includegraphics[width=1\columnwidth]{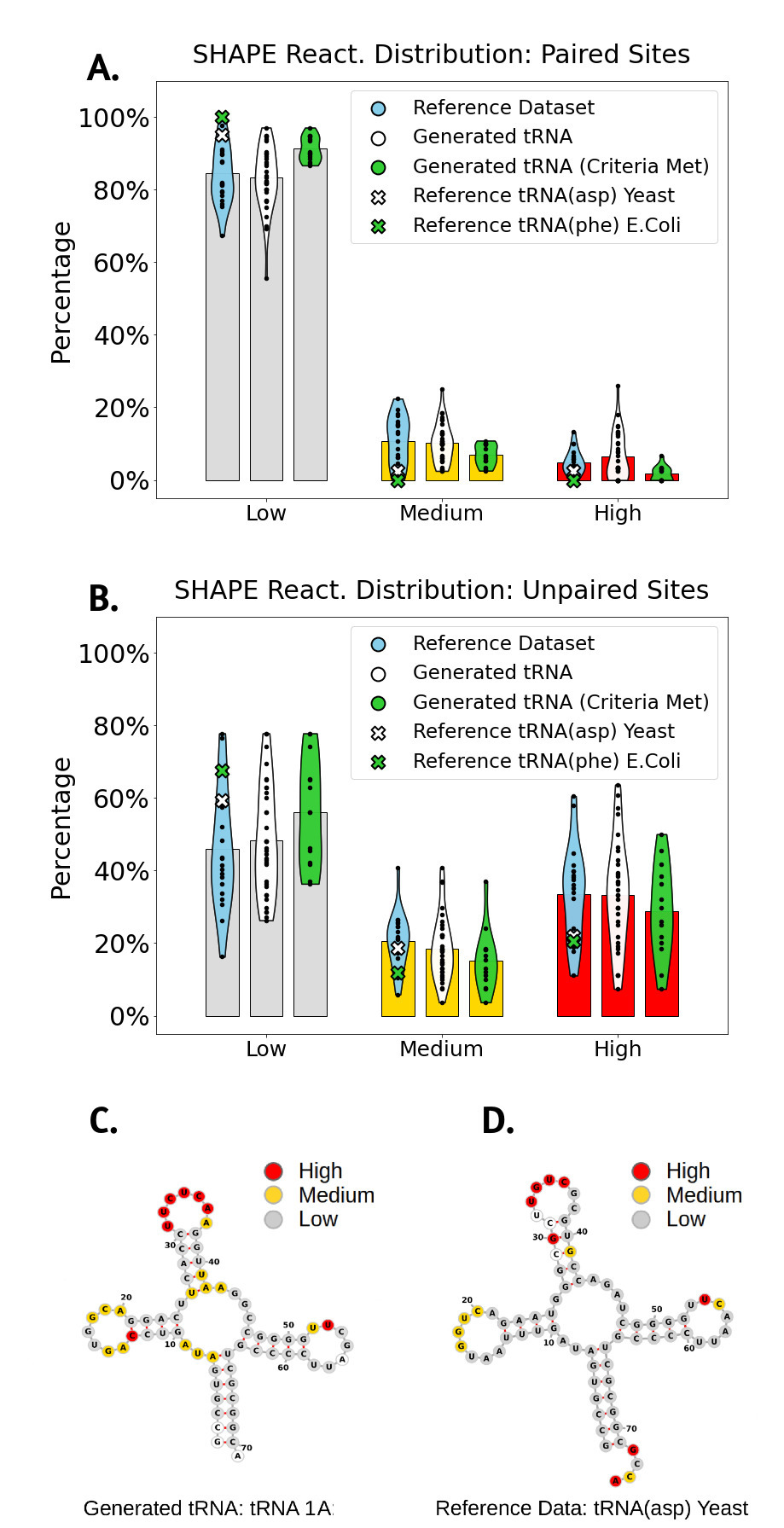}
\end{center}
\caption{\textbf{A.} Reactivity distribution for non paired residues, the bars refers to the indicated set average ('Reference SHAPE Dataset' $N=21$, 'Generated tRNA' $N=34$, 'Generated tRNA Criteria Me)' $N=14$ ).\textbf{B.} Reactivity distribution for paired residues. \textbf{C.} Example of reactivity-structure projection for the 1A molecule of the $14$ 'Generated tRNA (Criteria Met)'. \textbf{D.} Example of reactivity-structure projection for 'Reference SHAPE Data' tRNA(asp) Yeast}
\vspace{0mm}
\label{SHAPE}
\end{figure}

The definite test for the generative capacity of a statistical models of biomolecular sequences are experiments performed on artificially sampled sequences. As a first step in this direction, we have performed SHAPE-MaP experiments, which provide non-trivial structural information: chemical probing reveals different reactivities for nucleotide positions, which are paired vs. unpaired in the secondary structure of the tested RNA molecule \citep{SHAPEint}. 
This information is statistical: in the Reference dataset of published SHAPE experiments (cf.~{\em Materials and Methods}), out of the paired sites typically more than 80\% have low, less than 10\% high reactivity, while in the unpaired sites less than 50\% have low and around 40\% have high reactivity, cf.~Fig.~\ref{SHAPE}. 
Determining the specific pairing status of individual pairs is non-trivial due to a number of confounding factors: first, the correlation between SHAPE reactivity and base pairing is nonlinear. Second, SHAPE data may not mirror a single structure, but an average reactivity across a structural ensemble. Third, SHAPE reactivity can also be influenced by factors beyond secondary structure, such as base stacking and tertiary contacts \citep{SHAPEint}. Nevertheless, SHAPE-MaP experiments are a valuable instrument for assessing whether the SHAPE profile of a tested RNA molecule -- natural or artificial -- is statistically coherent with its expected secondary structures.

We used the tRNA family (RF0005) discussed above for generating 76 artificial tRNA sequences. We probed the SHAPE reactivities at each site of these sequences. We categorized the reactivities into three classes: low, medium, and high (as is common practice \citep{RNAstructure,SHAPEint}) and we assessed the distribution of these classes among paired and unpaired residues for each sequence. Due to experimental reasons we did not sample the 76 sequences freely from $P(a_1,...,a_L)$, but we introduced two types of constraints:
\begin{itemize}
    \item Due to experimental constraints, the last 16 nucleotides were kept constant, cf.~{\em Materials and Methods} and SI. Only the first 55 positions were generated by the model conditioned to the last 16, i.e.~they were sampled from $P(a_1,...,a_{55}\,|\,a_{56},...,a_{71})$. 
This reduces the effective sequence space $\Omega = e^S$ from $\sim 10^{22}$ sequences (cf.~Table \ref{table}) to $\sim  10^{14}$ , which is still a huge number beyond the possibility of exhaustive testing.
    \item Inspired by works about proteins \citep{russetal,malbranke2021improving}, only sequences of low energy ($E<44$) and good secondary-structure score ($F >0.53$, measured as the F-score between the RNAfold (October 2022) \citep{ViennaRNA} predicted structure and the tRNA consensus one)  were included in the test, cf.~the details given in the SI. These filters come at relatively low cost: while the energy-based filter is met by about 50\% of all sampled sequences, the double filter still preserves about 20\% of the sequences, inducing thus a very moderate decrease of the size of the sequence space.
\end{itemize}
For a more detailed overview of the dataset used in the test, please refer to the SI. Finally after probing, 34 of the 76 tRNAs satisfied the experimental standard of possessing reactivity data for more than 50\% of the sequence positions and were included in our further analyses.

In Fig.~\ref{SHAPE}, we observe that the reference dataset and the generated tRNA behave similarly, with clearly visible differences between paired and unpaired sites. We employed Permutational Multivariate Analysis of Variance (PERMANOVA) to test for statistical differences between the reactivity distributions of the reference dataset and of the generated tRNA, and between paired and unpaired sites. While we do not see indications for statistically significant differences between the reference dataset and the generated sequences ($p$-values of 0.993 for paired sites, 0.420 for unpaired sites), the paired and unpaired sites in the generated sequences are significantly distinct ($p$-value $1.9\times 10^{-7}$). 

Moreover, observing that the statistics for paired residues are more rigorous, especially on the two `Reference SHAPE dataset' tRNA, we decided to implement an additional filtering criterion. We deem artificial molecules as `Criteria Met' if over $85\%$ of their paired residues fell into the low reactivity class.
$14$ out of $34$ generated tRNA are classified as `Criteria Met'. Those are also the sequences that better pass the qualitative visual criterion (Fig.~\ref{SHAPE}C and SI). 

These results, albeit rather qualitative, indicate that the SHAPE reactivities of our artificially generated tRNAs are as consistent with the desired tRNA secondary structure, as the sequences in the reference dataset are with their published secondary structures, thereby supporting the validity of the eaDCA model as a generative statistical model.

\section{Conclusion And Outlook}

As in many disciplines and thanks to the strong increase in data availability, generative models gain growing importance in the modeling of biomolecular sequences. A first practical reason is quite obvious: generative models are of high biotechnological interest in biomolecular optimization and {\em de novo} design, directly or in combination with screening or selection assays when suggesting functionally enriched sequence libraries.

A second reason is less obvious, but has the potential to be at the basis of a paradigmatic shift in computational molecular biology. Traditionally, sequence bioinformatics was dominated by simpler statistical models, like the profile or covariance models discussed also in this paper, and which are of great success in analyzing extant biomolecular sequences, detecting homology, annotating sequences functionally, establishing RNA or protein families, reconstructing their phylogenies or aligning sequences. Generative models have the potential to go substantially beyond this, and to substantially contribute to our future understanding of biological molecules in their full complexity as high-dimensional, disordered and interacting systems. When a model is capable to generate diversified but viable artificial sequences, it necessarily incorporates essential constraints, which are functionally or structurally imposed on the sequences in the course of evolution. Even in this case, there is no guarantee that only such essential constraints are present in the model, and that these are encoded in a biologically interpretable way. In our work, we are therefore searching for {\em parsimonious} models, which contain as few as possible useless constraints (by using an information-theoretic criterion for including constraints, or the corresponding parameters, into the modeling), and which in turn should be maximally interpretable.

However, generative modeling is not trivial. The total sequence space is enormous, while the example sequences in RNA or protein family databases are quite limited. Very different models may be generative. In a parallel effort, \citep{remisimona} proposed and experimentally validated restricted Boltzmann machines (RBM) as generative models. In the case of protein families, it was shown before that RBM, which are shallow latent-space models, are able to detect extended functional sequence motifs \citep{RBMmotif,shimagaki2019selection}, but at the same time they have difficulties in representing pairwise structural constraints like residue contacts. On the contrary, eaDCA was found to easily detect contacts, but the patterns responsible for the clustered structure of families into subfamilies, easily visible by dimensional-reduction techniques like principal component analysis, remain hidden in the coupling network. It remains a challenge for the future to combine such different approaches to further improve interpretability of generative models.

Another problem is that, by definition, generative models reproduce statistical features of the training data, but there is no guarantee that statistical similarity implies functionality - this dilemma is well known from text or image generation with generative models, which do not always produce correct text contents or possible images. However, generative modeling will naturally benefit from the parallel evolution of more and more quantitative high-throughput experimental approaches in biology. On one hand, these can be used naturally to test model predictions (e.g. mutational effects) and sequences generated by the models, going far beyond the low-throughput experiments we presented in this predominantly computational work. On the other hand, these techniques substantially change the data situation in biology in several aspects (cf.~e.g.~\citep{russetal,tRNAfitness}): while current dataset, i.e.~MSA of homologous RNA or protein families, consist of positive but experimentally non annotated data, experiments provide (i) quantitative functional annotations for thousands of sequences and (ii) negative examples for artificial non-functional sequences generated by imperfect methods like random mutagenesis or sampling from imperfect models learned from finite data. This change in data will trigger future methodological work to develop integrative methods using all biologically relevant available information within the modeling process.

\section{Acknowledgements}

We are grateful to Sabrina Cotogno, Matteo Bisardi, Roberto Netti and Vaitea Opuu for helpful discussions during the project and the writing of the paper. We acknowledge also funding by the Institut Pierre-Gilles de Gennes (ANR-10-EQPX-34, to PN), EU H2020 Grant ERC AbioEvo (101002075, to PN), Human Frontier Science Program (RGY0077/2019, to PN), EU H2020 grant MSCA-RISE InferNet (734439, to MW). \\
High-throughput sequencing was performed by the ICGex NGS platform of the Institut Curie supported by the grants ANR-10-EQPX-03 (Equipex) and ANR-10-INBS-09-08 (France Génomique Consortium) from the Agence Nationale de la Recherche ("Investissements d’Avenir" program), by the ITMO-Cancer Aviesan (Plan Cancer III) and by the SiRIC-Curie program (SiRIC Grant INCa-DGOS-465 and INCa-DGOS-Inserm-12554). Data management, quality control and primary analysis were performed by the Bioinformatics platform of the Institut Curie.

\section{Conflict Of Interest Statement} None declared.

\section{Data And Code Availability}
The data and the code used in this paper are available at
\hyperlink{github}{https://github.com/FrancescoCalvanese/FCSeqTools.jl/}

\bibliography{example}
\bibliographystyle{unsrtnat}

\setcounter{equation}{0}
\renewcommand{\theequation}{S\arabic{equation}}

\setcounter{figure}{0}
\renewcommand{\thefigure}{S\arabic{figure}}

\setcounter{table}{0}
\renewcommand{\thetable}{S\arabic{table}}

\setcounter{section}{0}
\renewcommand{\thesection}{S\arabic{section}}

\onecolumn

\begin{center}
    \huge\textbf{Supplementary Information }
\end{center}

\vspace{1cm}  

\section{eaDCA: Details and Analytical Derivations}

\subsection{Analytical Derivations}

In this section, we provide the analytical derivation of the iterative model updates. At each iteration, we aim at improving the generative model 
\begin{equation} \label{Potts}
\begin{gathered}
    P_{t}(a_1,\dots,a_L) = \frac{1}{Z_t}\exp\big\{-E_t(a_1,\dots,a_L)\big\} \\
    -E_t(a_1,\dots,a_L)=\sum_{i=1}^{L} h_i(a_i)+\sum_{(ij)\in  {\cal E}_t} J_{ij}(a_i,a_j)
\end{gathered}
\end{equation}
by changing a single coupling matrix between a single pair of positions, with the aim of maximizing the model log-likelihood given the training data $\mathcal{D} = (a_i^r \,|\, i = 1,\dots,L;\, r = 1,\dots,M)$. Here, ${\cal E}_t$ is the set of currently ``activated'' couplings, i.e.~the set of all position pairs currently connected by a non-zero coupling. The log-likelihood can be written as
\begin{equation} 
\log\mathcal{L}_t = \sum_{r=1}^M \omega_r \log P_t(a_1^r,\dots,a_L^r) ,
\end{equation}
with $\omega_r$ being the standard sequence reweighting explained in the main text, and usually applied in DCA. Using Eq.~\eqref{Potts} and the definition $M_{\rm eff}=\sum_r \omega_r$ of the effective sequence number, we have
\begin{equation} 
\begin{gathered}
\log\mathcal{L}_t = -M_{\rm eff}\log Z_t + \sum_{r=1}^M \omega_r
\bigg(\sum_{i=1}^{L} h_i(a^r_i)+\sum_{ij\in {\cal E}_t} J_{ij}(a^r_i,a^r_j) \bigg) \ .
\end{gathered}
\end{equation}
To obtain the model $P_{t+1}(a_1,\dots,a_L)$, we add a $\Delta J_{mn}(a,b)$ to the couplings of $-E_t(a_1,\dots,a_L)$ in Eq.~\eqref{Potts}; this is done for a single position pair $1\leq m < n \leq L$, but arbitrary entries $a,b\in\{A,C,G,U,-\}$:
\begin{equation} 
\begin{gathered}
\log\mathcal{L}_{t+1} = -M_{\rm eff}\log Z_{t+1} + \sum_{r=1}^M \omega_r
\bigg(\sum_{i=1}^{L} h_i(a^r_i)+\sum_{ij\in  {\cal E}_t} J_{ij}(a^r_i,a^r_j) + \Delta J_{mn}(a^r_m,a^r_n)\bigg) 
\end{gathered}
\end{equation}
We aim at finding the $\Delta J_{mn}$ maximizing the likelihood gain
\begin{equation} 
\begin{gathered}
\Delta \log\mathcal{L} = \log\mathcal{L}_{t+1} -\log\mathcal{L}_t= -M_{\rm eff}\log \frac{Z_{t+1}}{Z_t}  
+ \sum_{r=1}^M \omega_r \Delta J_{mn}(a^r_m,a^r_n) \ .
\end{gathered}
\end{equation}
This can be simplified using the empirical two-point frequencies 
$f_{mn}(a,b) = \frac{1}{M_{\rm eff}} \sum_r \omega_r \delta_{a,a_m^r} \delta_{b,a_n^r}$:
\begin{equation}  \label{Delta}
\begin{gathered}
\frac{\Delta \log\mathcal{L}}{M_{\rm eff}} =  -\log \frac{Z_{t+1}}{Z_t}  + \sum_{a,b}  \Delta J_{mn}(a,b)f_{mn}(a,b) \ .
\end{gathered}
\end{equation}
The ratio $\frac{Z_{t+1}}{Z_t}$ of the partition functions can be expressed as
\begin{equation} 
\frac{Z_{t+1}}{Z_t} = \frac{\sum_{a_1, a_2, \ldots, a_L} \exp\big\{-E_t(a_1,\dots,a_L)\big\} \cdot\exp\big\{ \Delta J_{mn}(a_m, a_n)\big\}}{\sum_{a_1, a_2, \ldots, a_L} \exp\big\{-E_t(a_1, a_2, \ldots, a_L)\big\}} \ ,
\end{equation}
resulting in a simple mean value over $P_t$
\begin{equation}\label{rapporto}
    \frac{Z_{t+1}}{Z_t} = \left\langle e^{\Delta J_{mn}(a_m, a_n)} \right\rangle_{P_t}
    = \sum_{a, b} e^{\Delta J_{mn}(a, b)} P^t_{mn}(a, b)\ ,
\end{equation}
with $P^t_{mn}(a,b)$ being the two-site marginal of model $P_t$ for positions $m,n$. Substituting the last expression back into Eq.~\eqref{Delta}, we can express the likelihood change explicitly in terms of the changed coupling and the marginal probability of the old model at iteration $t$:
 \begin{equation}
     \frac{\Delta \log\mathcal{L}}{M_{\rm eff}} = -\log \left( \sum_{a, b} e^{\Delta J_{mn}(a, b)} P^t_{mn}(a, b) \right)   + \sum_{a,b}  \Delta J_{mn}(a,b) f_{mn}(a,b) 
 \end{equation}
For any given pair $(m,n)$, we can maximize this expression over $\Delta J_{mn}(a,b)$ by solving
\begin{equation}
  0 = f_{mn}(a,b) - \frac{e^{\Delta J_{mn}^*(a, b)} P^t_{mn}(a, b)}{\sum_{c, d} e^{\Delta J_{mn}^*(c,d)} P^t_{mn}(c,d)} \ ,
\end{equation}
i.e.~by choosing
\begin{equation}
\Delta J^*_{mn}(a, b) = \log \left( \frac{f_{mn}(a, b)}{P^t_{mn}(a, b)} \right) \ .
\end{equation}
For given $(m,n)$, the maximally realizable likelihood gain therefore reads
\begin{equation}
    \frac{\Delta \log\mathcal{L}}{M_{\rm eff}} 
    = \sum_{a,b } f_{mn}(a,b) \log \bigg( \frac{f_{mn}(a, b)}{P^t_{mn}(a, b)} \bigg) = D_{KL}\left(f_{mn} \,||  \, P^t_{mn}\right) \ .
\end{equation}
To finalize the derivation of our algorithm we need to select, out of all possible position pairs, the one realizing the largest likelihood gain,
\begin{equation} \label{DKL}
      (kl) = \operatorname*{argmax}_{1\leq m < n \leq L} \,D_{KL}\left(f_{mn} \,||  \, P^t_{mn}\right) \ ,
\end{equation}
and update only the corresponding coupling 
\begin{equation} \label{update}
    \Delta J^*_{kl}(a,b) = \log \left( \frac{f_{kl}(a,b)}{P^t_{kl}(a,b)}\right)
\end{equation}
for all $a,b\in\{A,C,G,U,-\}$. Note that Eq.~\eqref{DKL} optimizes over all pairs of positions, including those already present in ${\cal E}_t$. 
This choice results in the following change in the Potts Model:
\begin{eqnarray}
    E_{t+1}(a_1,...,a_L) &=& E_{t}(a_1,...,a_L) - \Delta J_{kl}^*(a_k,a_l) \ ,
    \nonumber\\
    {\cal E}_{t+1} &=& {\cal E}_t \cup \{(kl)\} \ .
\end{eqnarray}

\subsection{Monte Carlo Sampling} \label{MCMC}

Due to the high computational cost of calculating the exact two-point  probabilities $P^t_{kl}(a,b)$ for the coupling update in Eq.~\eqref{update}, we resort to a Monte Carlo approach to obtain an approximation. The Monte Carlo simulation involves sampling a number of states (sequences) from $P_t(a_1, a_2, \ldots , a_L)$, and using the sample two-point frequencies $P^{m_1}_{kl}(a,b)$ as an estimate for the model two-point probabilities $P^t_{kl}(a,b)$. 
Our chosen Monte Carlo algorithm is Gibbs sampling. The default implementation use $10000$ independent chains to compute $P^{m_1}_{kl}(a,b)$. To optimize computational time, we used the Persistent Contrastive Divergence technique in our implementation. Essentially, this approach initializes the Monte Carlo runs for the sampling of $P_t(a_1,a_2,\dots,a_L)$ to samples obtained from the $P_{t-1}(a_1,a_2,\dots,a_L)$ at the previous step. Given that two consecutive models differ by just one interaction coupling, the samples from $P_{t-1}$ are already close to equilibrium for $P_{t}$. The default implementation performs $5$ Gibbs sweeps on all the $10000$ training chains. 
 Upon selecting our final generative model, we performed an independent re-sampling starting from randomized sequences. This step ensured that the Persistent Contrastive Divergence did not trap the simulation in a locally stable state.

\subsection{Regularization} \label{REG}

Suppose that in our Monte Carlo we sample a sequence with a pair of nucleotides at sites $(k,l)$ that is absent from the natural dataset. In this case, if the eaDCA traverses the edge in question, the resulting coupling update Eq.~\eqref{update} is $-\infty$. Similarly, if a sampled pair present in the natural dataset is never sampled, we encounter a $+\infty$ coupling update. 
These conditions, and in general  excessively high valued couplings, pose threats to the performance and to the ergodicity of the models.

To mitigate these issues, we introduce a regularization for the update:
\begin{equation} \label{real_update}
\Delta J^*_{kl}(a,b) = \log\left(\frac{(1-\alpha)f_{kl}(a,b)+\frac{\alpha}{q^2}}{(1-\alpha)P^t_{kl}(a,b)+\frac{\alpha}{q^2}}\right)
\end{equation}

As the value of $\alpha$ increases, the regularization effect on the update correspondingly intensifies. In our models, we used a fixed $\alpha = 0.1$. This choice allowed our model to pass the ergodicity tests in the re-sampling and achieve high performance scores. 

Interestingly, despite this modification, we note from Eq.~\eqref{real_update} that the termination condition remains the same -- the equality between the natural two-point natural frequencies $f_{kl}(a,b)$ and the model two-point probabilities $P_{kl}(a,b)$. 

\subsection{Partition Function Conservation}

In this section we provide the proof that eaDCA preserves the partition function $Z$. 
Substituting the coupling update Eq.~\eqref{update} in Eq.~\eqref{rapporto} we can express the ratio $\frac{Z_{t+1}}{Z_t}$ as:
\begin{equation} \label{Zconv}
\frac{Z_{t+1}}{Z_t} = \sum_{a, b} \exp\bigg\{\log\bigg(\frac{f_{kl}(a,b)}{P^t_{kl}(a,b)}\bigg)\bigg\} P^t_{kl}(a, b)
 = \sum_{a, b}\frac{f_{kl}(a,b)}{P^t_{kl}(a,b)} P^t_{kl}(a, b)
 = \sum_{a, b} f_{kl}(a,b)
=1 \ ,
\end{equation}
which proves that $Z_{t+1}=Z_t$ . Since our procedure starts from the Profile Model, which has a $Z_0=1$, all the models in the eaDCA chain are normalized. A nice consequence is that the models entropy are easily accessible trough the relation
\begin{equation}
S_t  =<E_t>_{P_t}
\end{equation}
In Section~\ref{MCMC} and Section~\ref{REG} we discussed the impossibility of applying the ideal algorithm. This means that the real coupling update deviates from Eq.~\eqref{update}. Its real value is:
\begin{equation} \label{real_update}
    \Delta J^*_{kl}(a,b) = \log\bigg(\frac{(1-\alpha)f_{kl}(a,b)+\frac{\alpha}{q^2}}{1-\alpha)P^{m_1}_{kl}(a,b)+\frac{\alpha}{q^2}}\bigg)
\end{equation}
which takes in account both the regularization term $\alpha$ and the Monte Carlo estimation $P^{m_1}_{kl}(a,b)$.
This implies that Eq.~\eqref{Zconv} only holds approximately. 

Using Eq.~\eqref{rapporto} and Eq.~\eqref{real_update} we can get the real relation between $Z_{t+1}$ and $Z_t$ :
\begin{equation} 
Z_{t+1  }={Z_t}\sum_{a, b}\frac{(1-\alpha)f_{kl}(a,b)+\frac{\alpha}{q^2}}{(1-\alpha)P^{m_1}_{kl}(a,b)+\frac{\alpha}{q^2}} P^t_{kl}(a, b) \ ,
\end{equation}
or
\begin{equation} \label{Iter}
\log Z_{t+1  }=\log {Z_t} + \log\bigg(\sum_{a, b}\frac{(1-\alpha)f_{kl}(a,b)+\frac{\alpha}{q^2}}{(1-\alpha)P^{m_1}_{kl}(a,b)+\frac{\alpha}{q^2}} P^t_{kl}(a, b) \bigg) \ .
\end{equation}
Hence, defining
\begin{equation} 
\Phi_t = \log\bigg(\sum_{a, b}\frac{(1-\alpha)f_{kl}(a,b)+\frac{\alpha}{q^2}}{(1-\alpha)P^{m_1}_{kl}(a,b)+\frac{\alpha}{q^2}} P^t_{kl}(a, b) \bigg) \ ,
\end{equation}
we can iterate Eq.~\eqref{Iter} to obtain
\begin{equation}  \label{logZ}
\log Z_{t}=\sum_{s=0}^{t-1} \Phi_s \ .
\end{equation}
However, we then are back to the initial problem: in order
to compute $\Phi_s$ with Eq.~\eqref{logZ}, we need the exact values of $ P^t_{kl}(a, b)$. A possible solution is to use another
independent Monte Carlo approximation $P^{m_2}_{kl}(a, b)$ to obtain
 \begin{equation} 
\Phi_t \simeq \log\bigg(\sum_{a, b}\frac{(1-\alpha)f_{kl}(a,b)+\frac{\alpha}{q^2}}{(1-\alpha)P^{m_1}_{kl}(a,b)+\frac{\alpha}{q^2}} P^{m_2}_{kl}(a, b) \bigg) \ .
\end{equation}
We tested how well these arguments hold using a very short RNA segment (first $11$
residues from the RF0442 family):
\begin{table}[htb]
    \centering
    \begin{tabular}{|c|c|c|c|c|} 
    \toprule
   $P^{m_1}_{kl}(a,b)$ (training) sequences  & $P^{m_2}_{kl}(a, b)$ ($\Phi_t$) sequences & eaDCA iterations & Exact $S$ & Estimated $S$ \\
    \midrule
   $1000$  & $2000$  & $390$  & $11.56$ & $11.72$ \\
    $1000$ & $10000$ & $390$ & $11.65$ & $11.65$ \\
    $1000$ & $2000$  & $1000$ & $11.41$  & $11.58$ \\
    $1000$ & $10000$  & $1000$ & $11.66$  & $11.66$ \\
    \bottomrule
    \end{tabular}
    \caption{We illustrate the entropy estimation using $2000$ and $10000$ chains to compute $P^{m_2}_{kl}(a, b)$. A reduced sample of $1000$ sequences, as opposed to the standard $10000$, was employed for training, enabling us to examine scenarios where the violation of the conservation of $Z$ is more pronounced.}
    \label{tab:my_label}
\end{table}

Given that this procedure results in a reasonable approximation of $\log Z_t$, we can still derive the models entropy using
\begin{equation} 
S_t =<E_t>_{P_t} + \log Z_{t }
\end{equation}
In the default eaDCA implementation we use $2000$ independent chains to compute $P^{m_2}_{kl}(a, b)$. 

\subsection{Typical eaDCA Training Process}

\begin{figure}[H]
  \centering
     \includegraphics[width=0.8\textwidth]{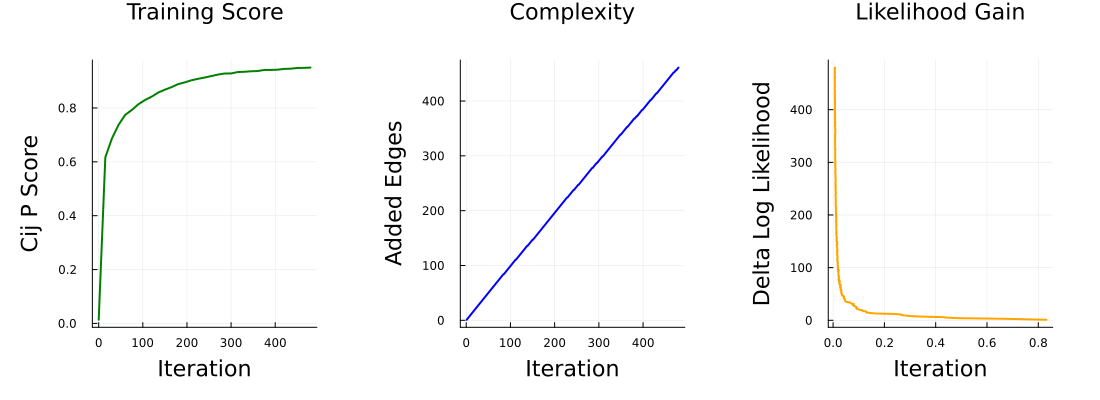} 
   \caption{eaDCA Training Process for the RF01734 Family. The plot showcases typical behavior of the $c_{ij}$ Pearson score, the number of added edges and the likelihood gain as the number of iterations increases.    
   }     
   \label{fig:training}
\end{figure}

In Fig.~\ref{fig:training} we report the typical training trajectory of eaDCA.
The first few iterations yield the highest gain in Pearson Score and likelihood. After this initial spike, both the likelihood gain and the increase in score start to plateau, becoming much slower. However, delving into this slower region is essential for obtaining models with good statistical performance. Complexity, on the other hand, increases almost linearly as the number of iterations rises. While coupling-update iterations do occur, they are relatively rare, allowing us to obtain a well-performing model before they become predominant. 

\section{Training Data And Statistical Tests}

\begin{table}[H]
    \centering
    
    \begin{tabular}{|c|c|c|c|c|c|c|c|} 
    \toprule
   Family Name & $L$ & $M$  & $M_{eff}$ & PR$\%$  & $S$ & $\Omega$ & PDB\\
    \midrule
    RF00005 & $71$  & $28770$   & $2267$   & $84.35\%$  & $51.34$   & $1.98\times 10^{22}$ & 1ehz A\\
    RF00028 & $251$ & $611$     & $233$    & $88.17\%$  & $106.31$  & $1.48\times 10^{46}$ & 1gid A Err\\
    RF00050 & $140$ & $4051$    & $530$    & $92.16\%$  & $73.78$   &  $1.10\times 10^{32}$ & 3f2q X\\
    RF00059 & $105$ & $12208$   & $4683$   & $94.34\%$  & $77.41$   & $4.16\times 10^{33}$ &  3d2g A \\
    RF00080 & $175$ & $468$     & $371$    & $64.41\%$  & $82.29$   & $5.47\times 10^{35}$ & 6cb3 A \\
    RF00114 & $117$ & $620$     & $218$    & $73.37\%$  & $58.72$   & $3.18\times 10^{25}$ & 2vaz A Err\\
    RF00162 & $108$ & $5974$    & $1283$   & $91.09\%$  & $66.23$   & $5.80\times 10^{28}$ & 3gx5 A\\
    RF00167 & $102$ & $2613$    & $1596$   & $90.18\%$  & $71.87$   & $1.63\times 10^{31}$ & 4tzx X \\
    RF00168 & $183$ & $2191$    & $1419$   & $82.42\%$  & $117.55$  & $1.13\times 10^{51}$ & 3dil A\\
    RF00169 &  $97$ & $4641$    & $796$    & $82.50\%$  & $64.67$   & $1.22\times 10^{28}$ & 1z43 A Err\\
    RF00234 & $171$ & $893$     & $639$    & $82.83\%$  & $95.86$   & $4.28\times 10^{41}$ & 2h0s B Err\\
    RF00379 & $136$ & $3808$    & $1428$   & $87.83\%$  & $89.85$   & $1.05\times 10^{39}$ & 4qln A \\
    RF00380 & $169$ & $1057$    & $149$    & $95.15\%$  & $84.24$   & $3.85\times 10^{35}$ & 3pdr A\\
    RF00442 & $108$ & $845$     & $312$    & $89.89\%$  & $56.02$   & $2.13\times 10^{24}$ & 5u3g B\\
    RF00504 &  $94$ & $4395$    & $1363$   & $91.19\%$  & $63.86$   & $5.42\times 10^{27}$ & 3ox0 B \\
    RF01051 &  $87$ & $4032$    & $1850$   & $90.24\%$  & $57.46$   & $9.01\times 10^{24}$ & 4yaz A\\
    RF01725 & $101$ & $707$     & $249$    & $80.32\%$  & $57.00$   & $5.69\times 10^{24}$ & 4l81 A\\
    RF01734 &  $65$ & $2345$    & $923$    & $77.84\%$  & $44.31$   & $1.75\times10^{19}$ & 4enc A Err\\
    RF01750 & $101$ & $1402$   & $420$    & $85.45\%$  & $59.20$ & $5.13\times10^{25}$ & 4xwf A \\
    RF01786 &  $85$ & $585$    & $336$    & $70.39\%$  & $48.40$   & $1.05\times 10^{21}$ & 5nwq A\\
    RF01831 & $100$ & $692$    & $233$    & $83.98\%$  & $48.64$   & $1.33\times 10^{21}$ & 4lvv A\\
    RF01852 &  $91$ & $2351$   & $296$    & $89.47\%$  & $52.13$   & $4.36\times 10^{22}$ & 3rg5 A \\
    RF01854 & $255$ & $1052$   & $343$    & $83.15\%$  & $92.02$   & $9.20\times 10^{39}$ & 4wfl A\\
    RF02001 & $171$ & $3103$   & $657$    & $70.21\%$  & $96.44$   & $7.64\times 10^{41}$ & 4y1o A\\
    RF02553 & $114$ & $194$    & $116$    & $81.54\%$  & $35.77$   & $3.43\times 10^{15}$ & 6cu1 A\\
    \bottomrule
    \end{tabular}
    \caption{Results from the application of eaDCA across all 25 test families are presented. The PDB identifiers are indicated, and if 'err' is displayed, it signifies an issue in mapping the PDB to the alignment.}
    \label{tab:my_label}
\end{table}

\begin{figure}[H]
  \centering
     \makebox[\textwidth]{\includegraphics[width=1\textwidth,height=0.45\textheight]{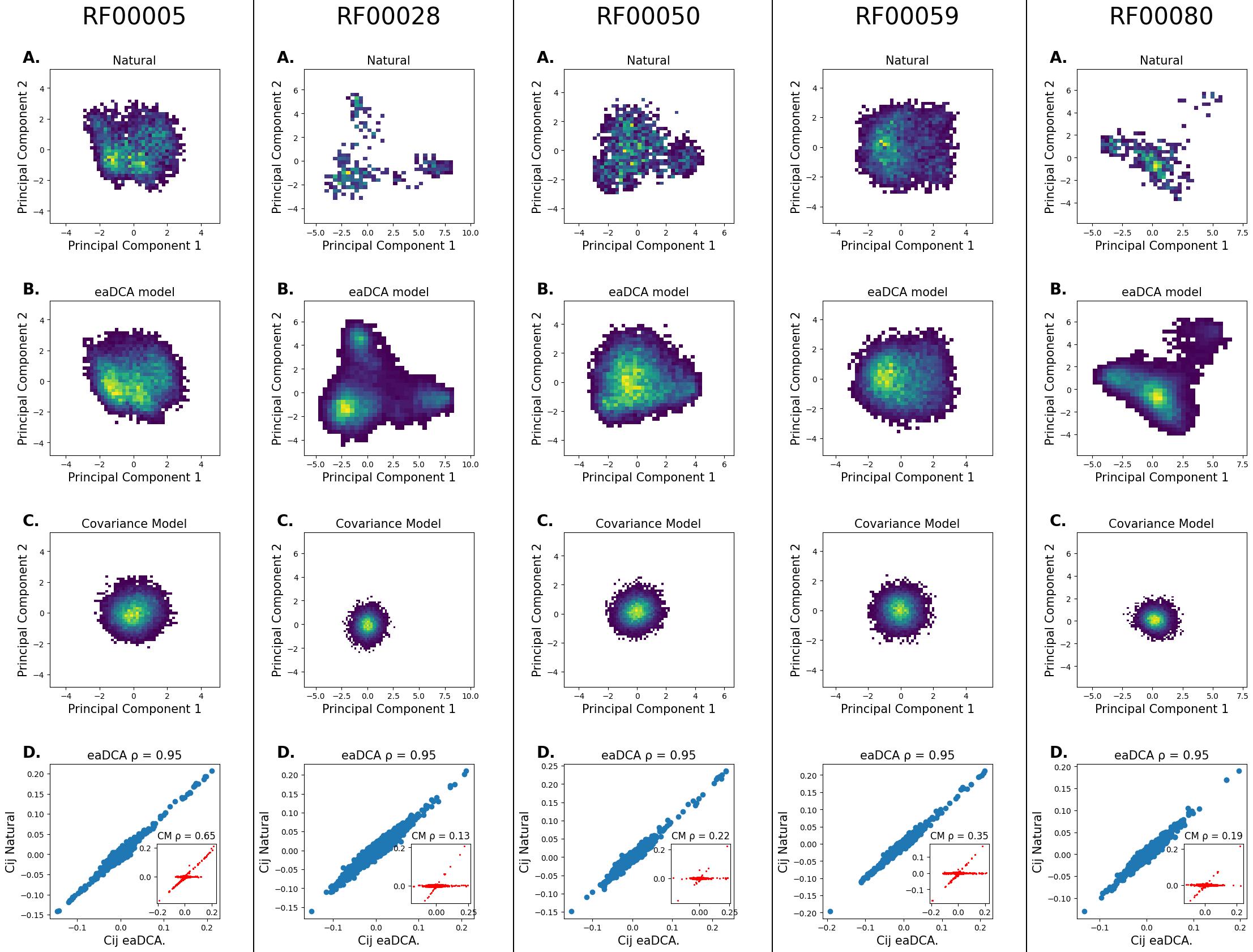}} 
\end{figure}
\begin{figure}[H]
  \centering
\makebox[\textwidth]{\includegraphics[width=1\textwidth,height=0.45\textheight]{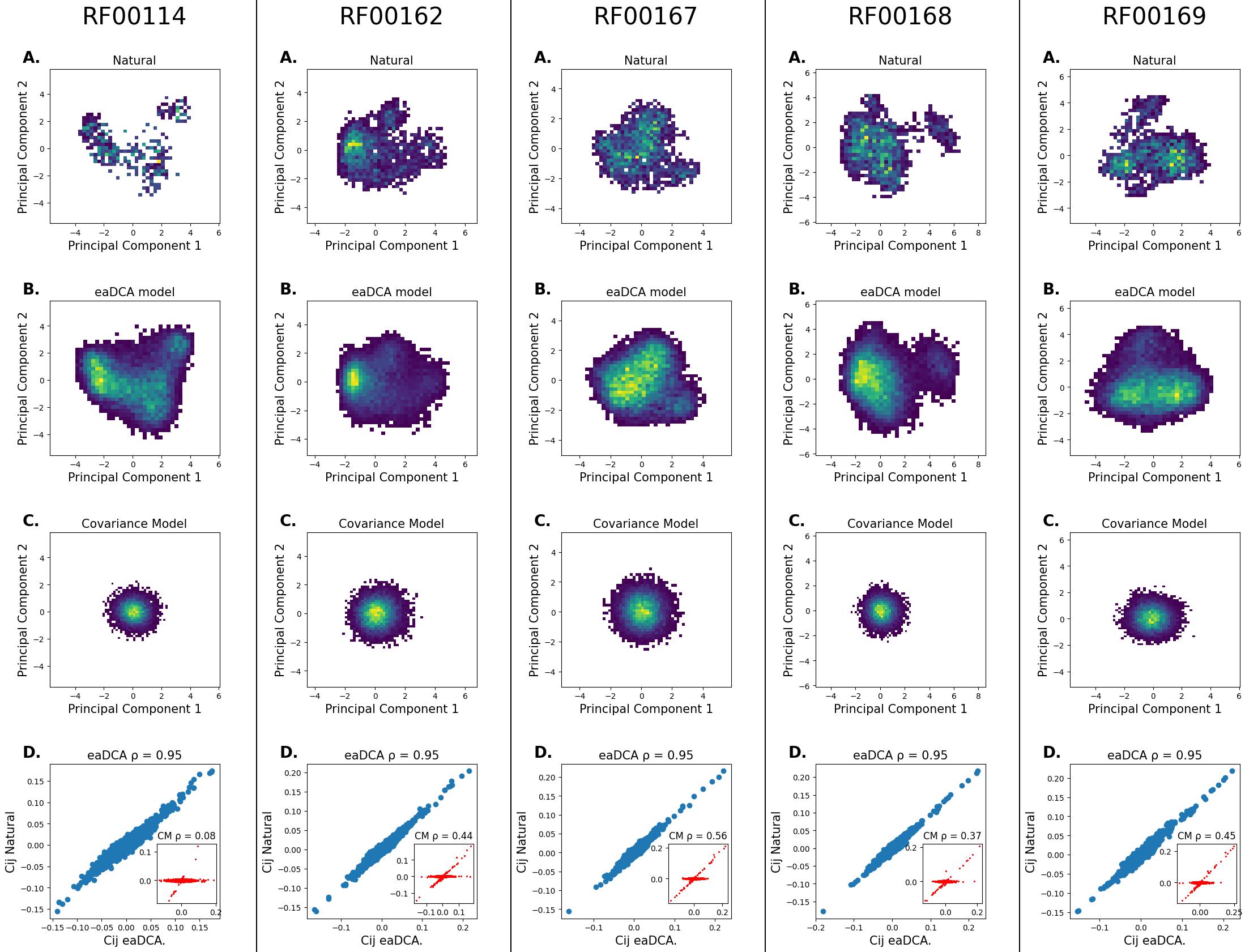}} 
\end{figure}

\begin{figure}[H]
  \centering
      \makebox[\textwidth]{\includegraphics[width=1\textwidth,height=0.45\textheight]{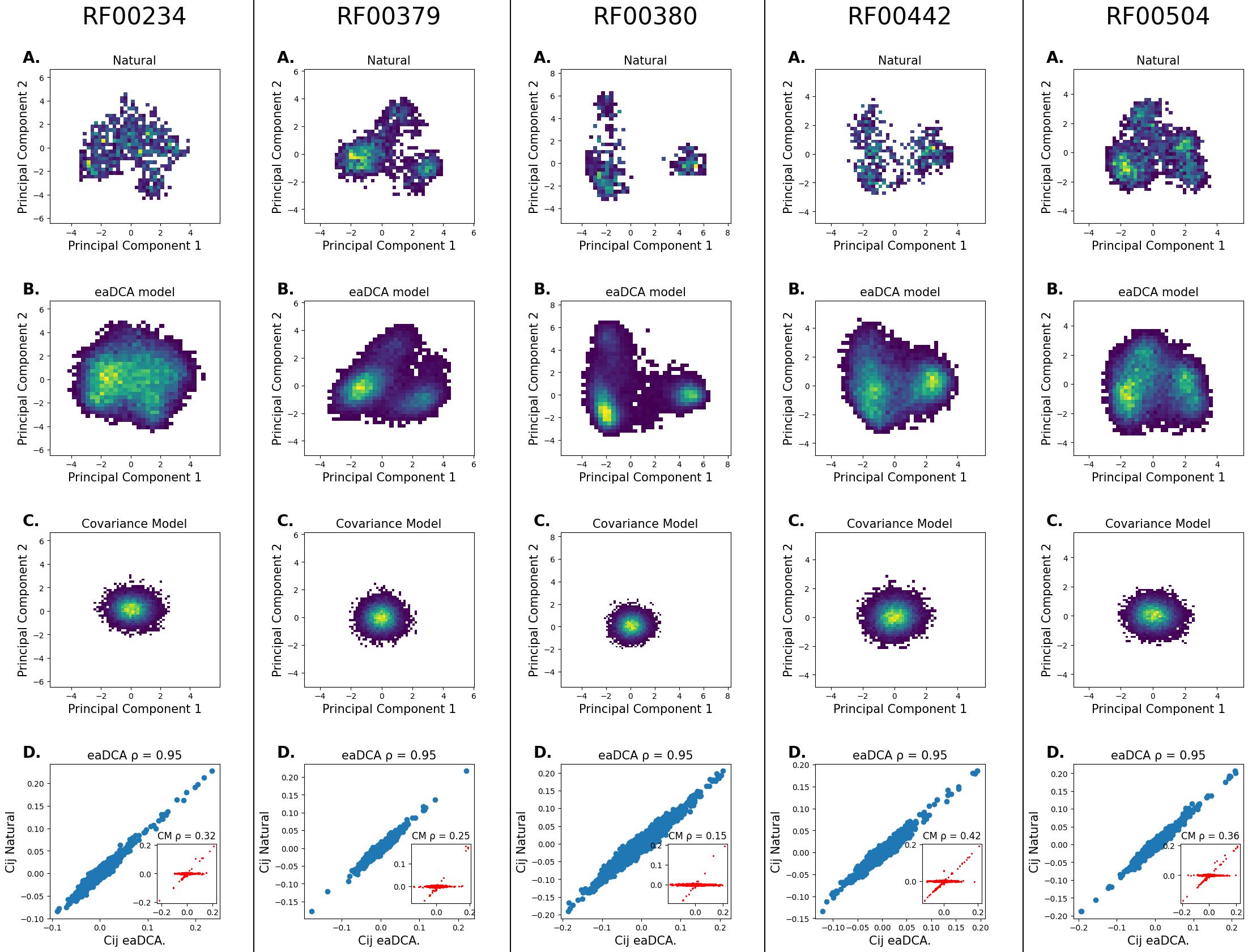}} 
\end{figure}
\begin{figure}[H]
  \centering
      \makebox[\textwidth]{\includegraphics[width=1\textwidth,height=0.45\textheight]{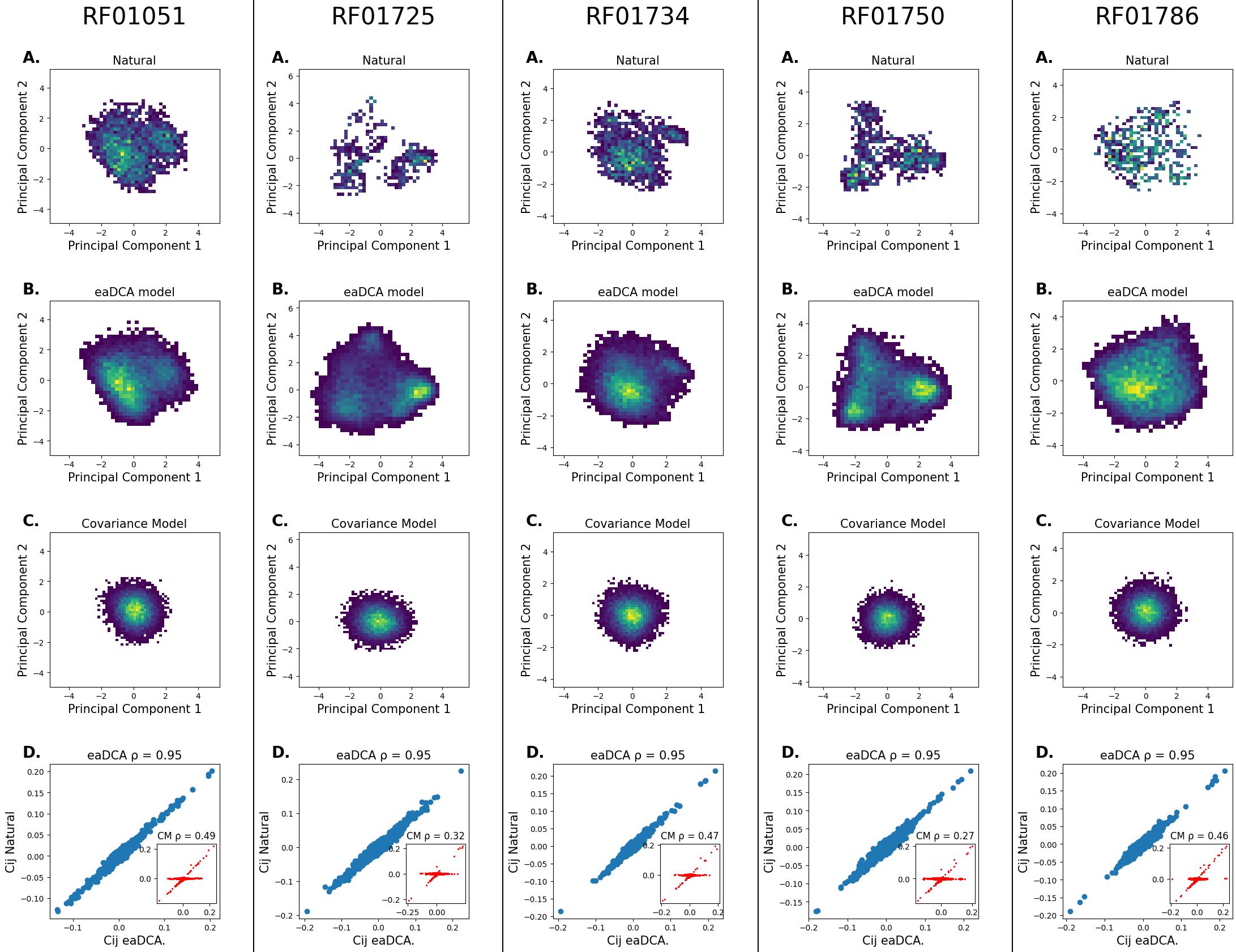}} 

\end{figure}

\begin{figure}[H]
  \centering
    \makebox[\textwidth]{\includegraphics[width=1\textwidth,height=0.45\textheight]{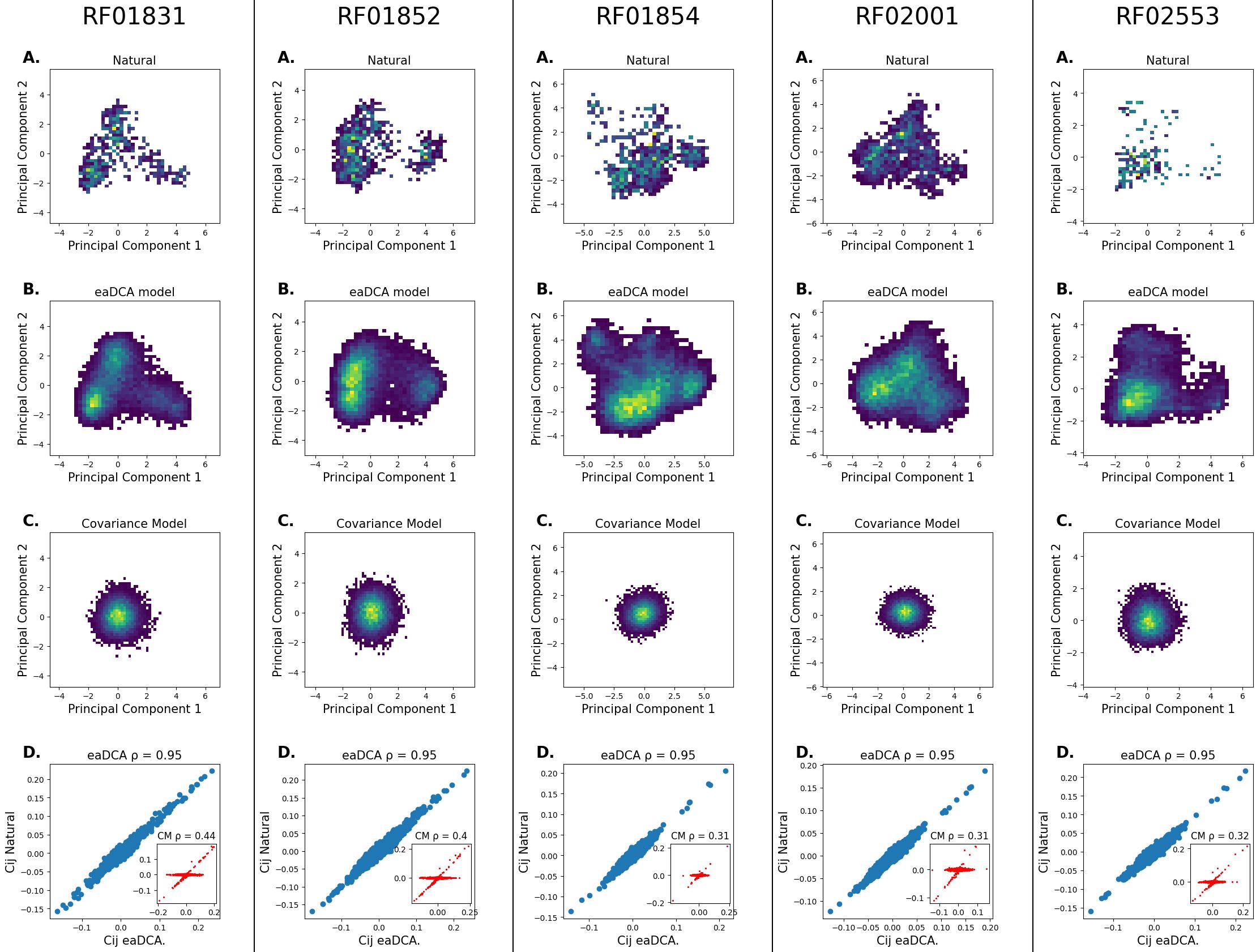}} 
       \caption{\textbf{A.} PCA dimensional reduction of natural sequences. \textbf{B.} PCA dimensional reduction of eaDCA generated sequences. \textbf{C.} PCA dimensional reduction of covariance model generated sequences.  \textbf{D.} two-point statistics representation for eaDCA model and covariance model.}
  \label{fig:Fitness Data Description}
\end{figure}

\section{Parameter Interpretation}
\begin{figure}[H]
  \centering
    \makebox[\textwidth]{\includegraphics[width=1.0\textwidth,height=0.85\textheight]{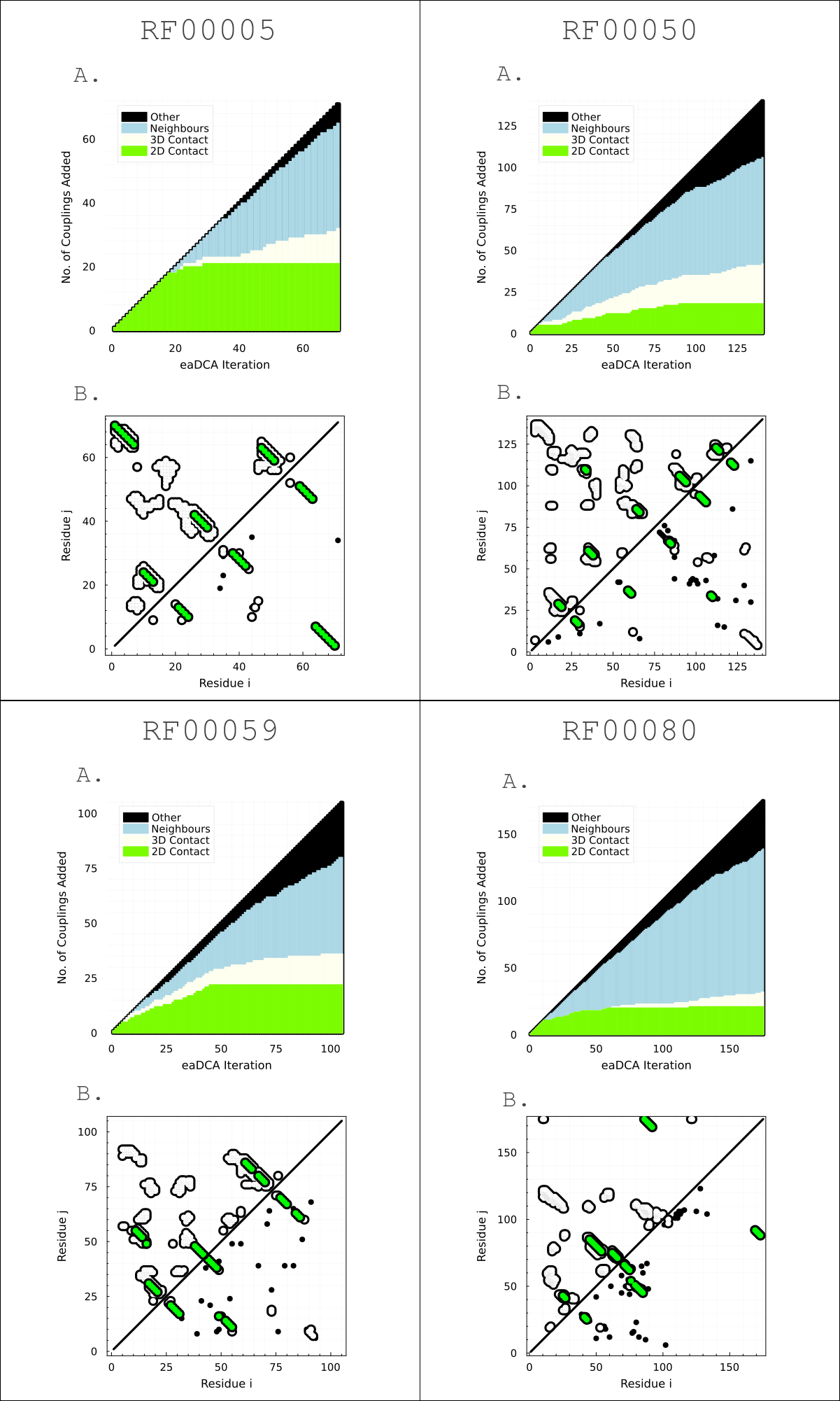}} 
\end{figure}
\begin{figure}[H]
  \centering
    \makebox[\textwidth]{\includegraphics[width=1.0\textwidth,height=0.85\textheight]{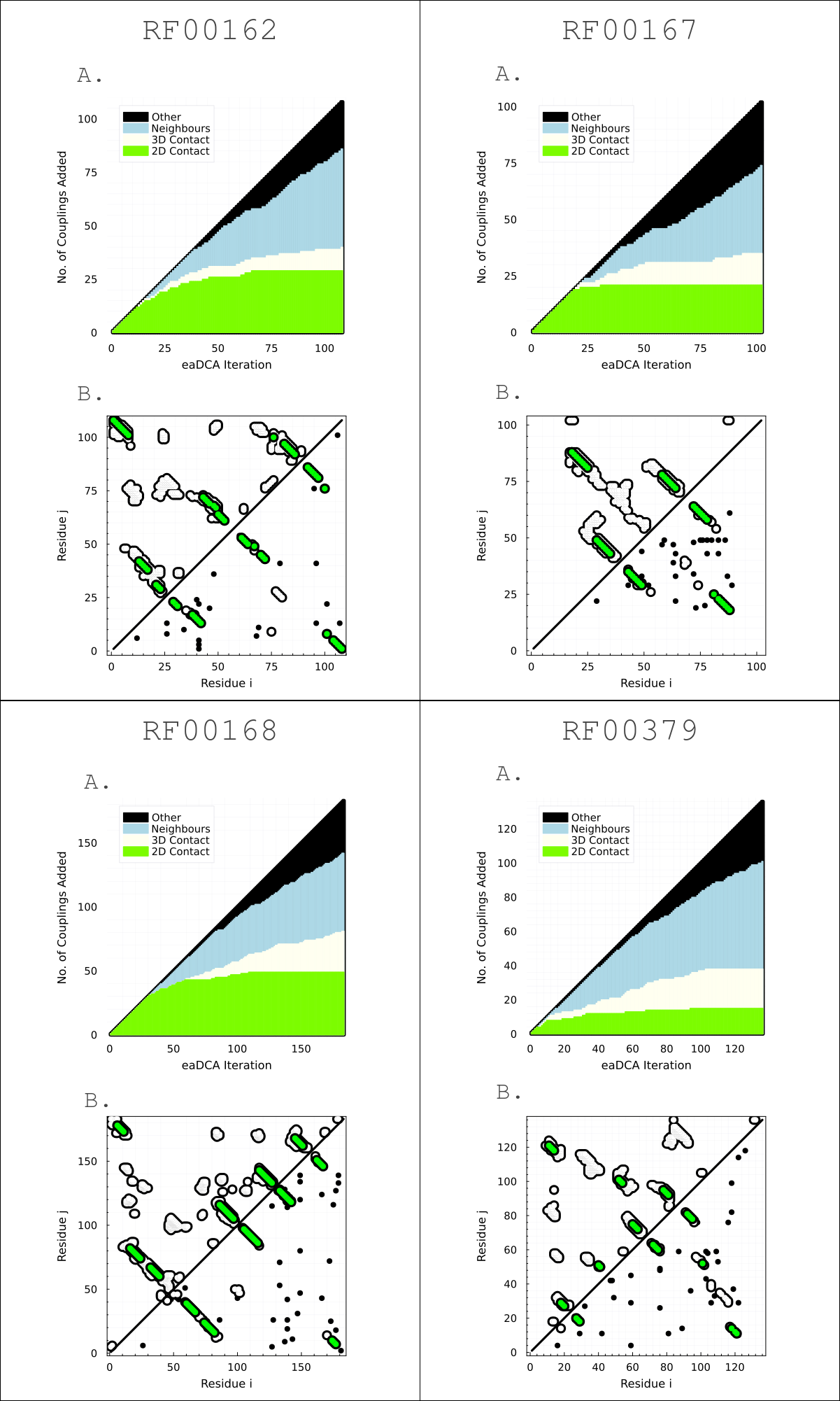}} 
\end{figure}
\begin{figure}[H]
  \centering
    \makebox[\textwidth]{\includegraphics[width=1.0\textwidth,height=0.85\textheight]{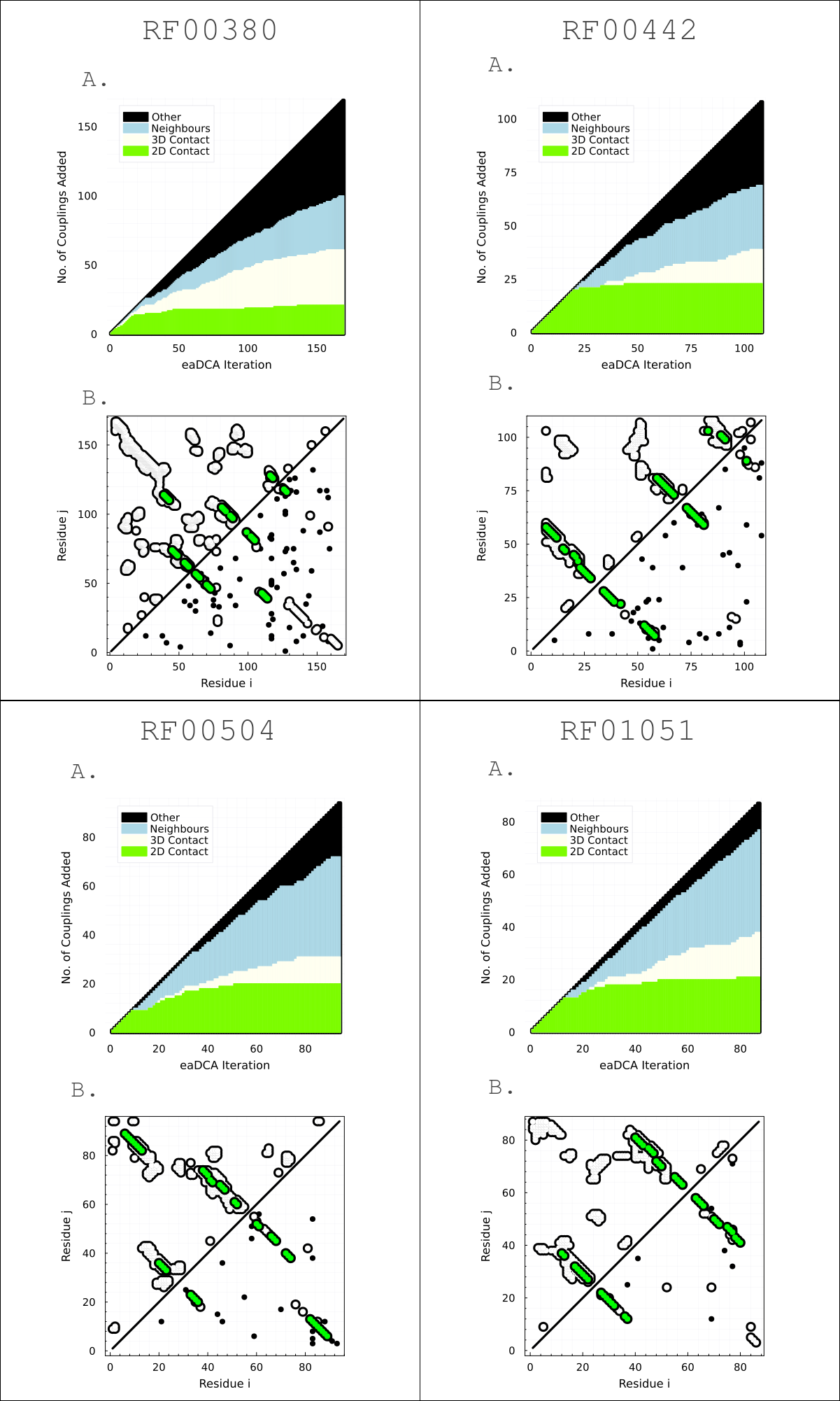}} 
\end{figure}
\begin{figure}[H]
  \centering
    \makebox[\textwidth]{\includegraphics[width=1.0\textwidth,height=0.85\textheight]{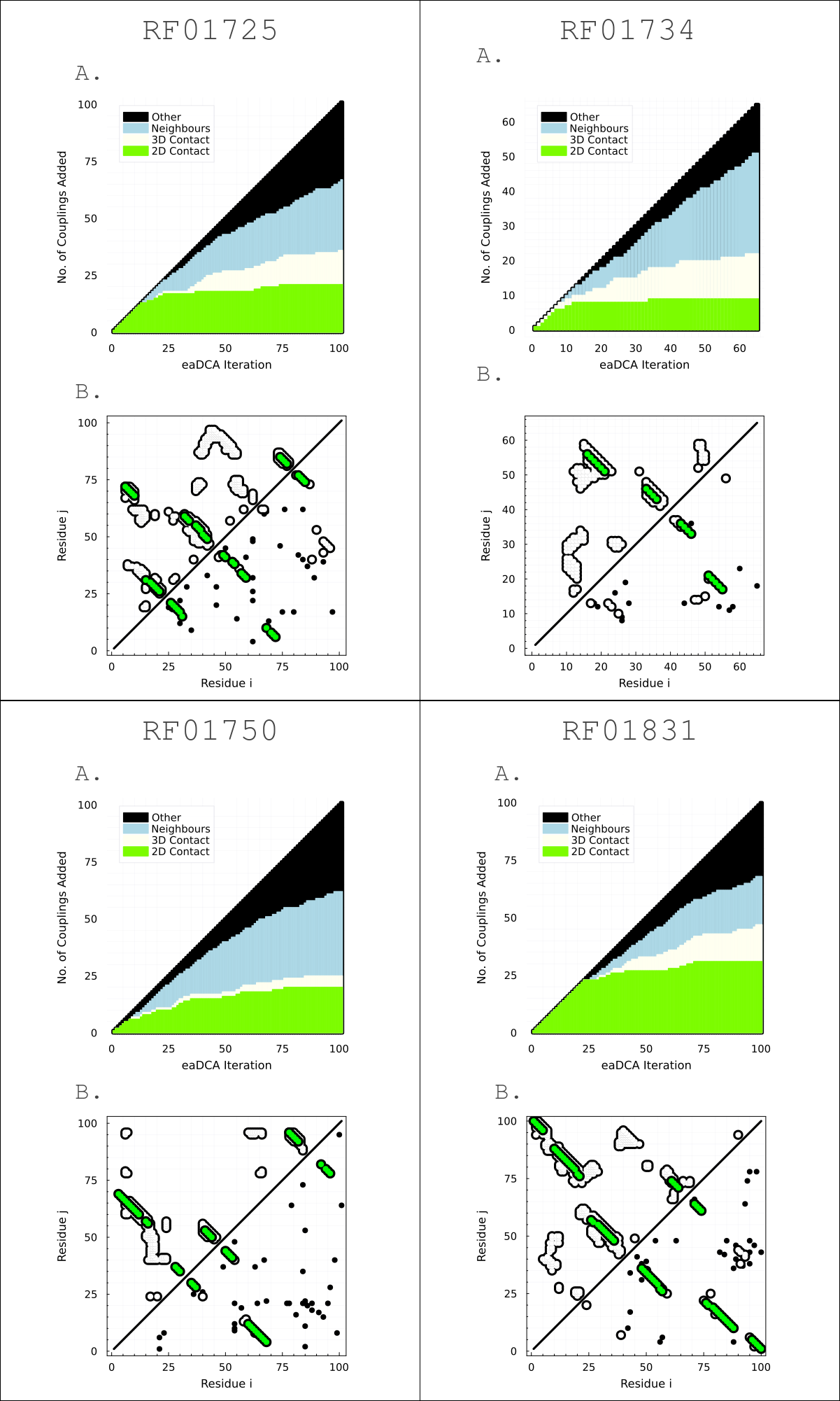}} 
\end{figure}
\begin{figure}[H]
  \centering
    \makebox[\textwidth]{\includegraphics[width=1.0\textwidth,height=0.85\textheight]{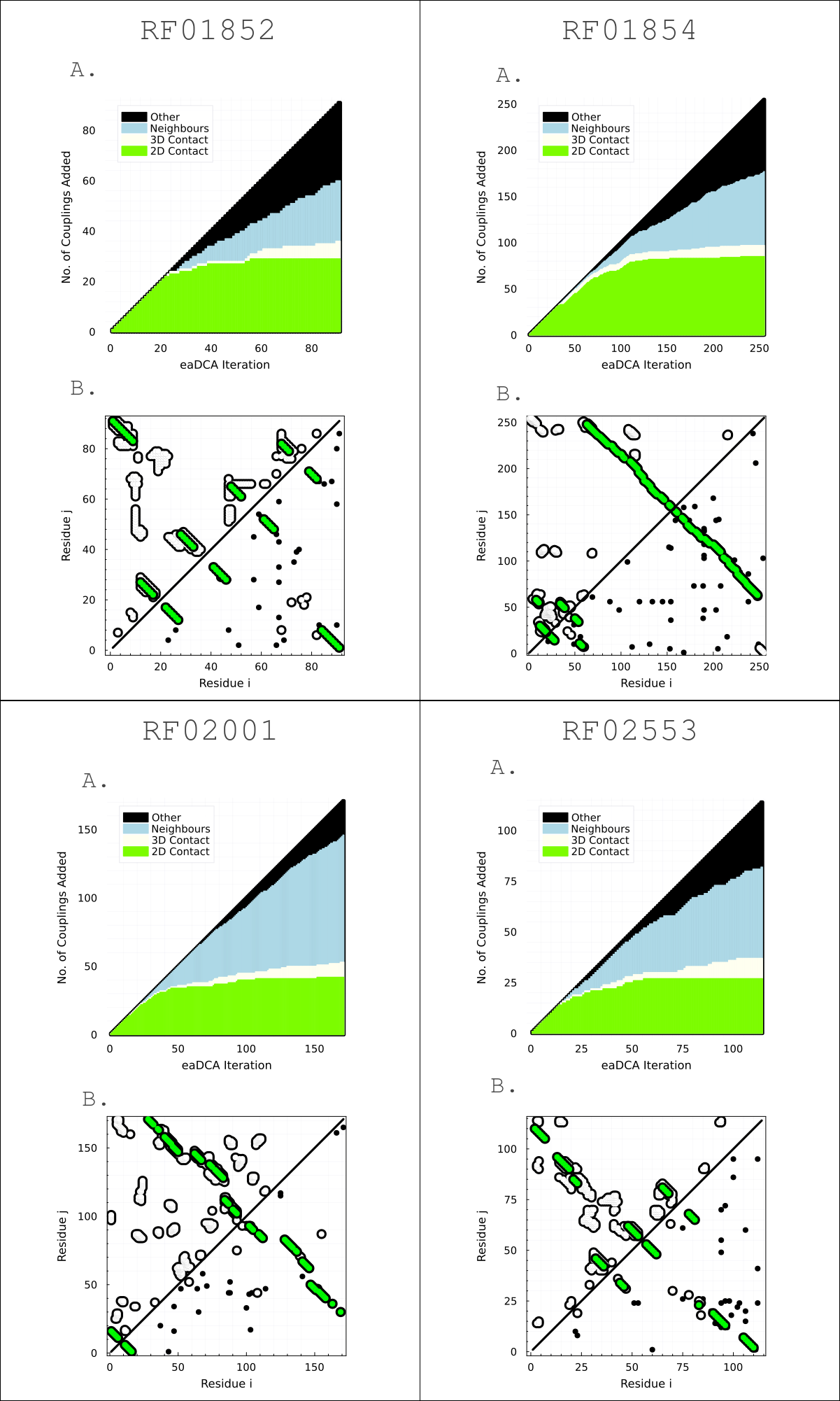}} 
 \caption{\textbf{A.} Firs $L$ edges classification. \textbf{B.} Contact-Map (upper-left) and added edges (lower-right). }
\end{figure}

\section{Prediction of Mutational Effects}
\subsection{Fitness Dataset Characteristics}
\begin{figure}[H]
  \centering
      \makebox[\textwidth]{\includegraphics[width=0.75\textwidth]{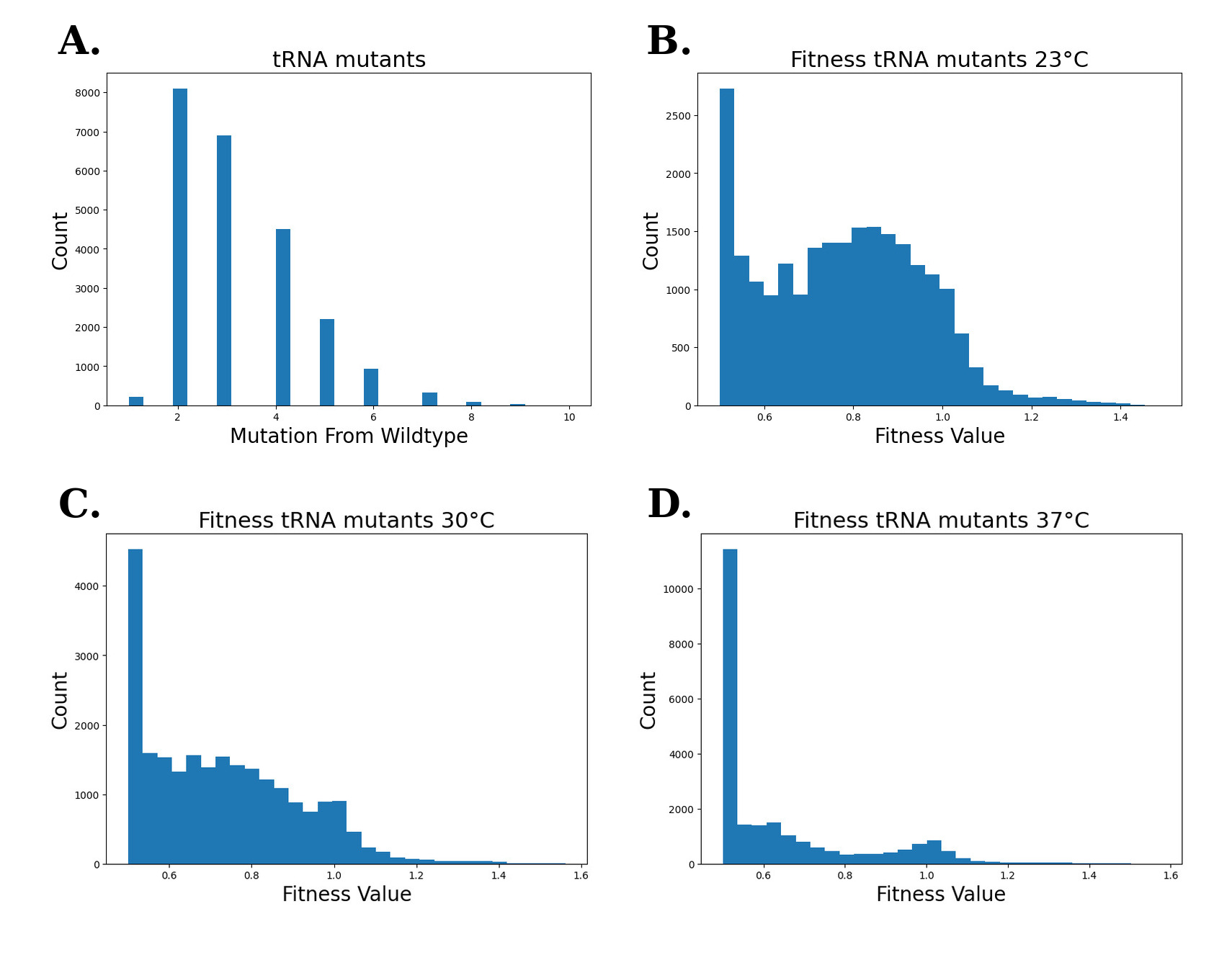}} 
  \caption{\textbf{A.} Distribution of the number of mutations from the wildtype. \textbf{B.} Fitness distribution at $23^{\circ}C$. \textbf{C.} Fitness distribution at $30^{\circ}C$. \textbf{D.} Fitness distribution at $37^{\circ}C$. }
  \label{fig:Fitness Data Descriptio}
\end{figure}

\subsection{Prediction of tRNA Mutational Effects at $\mathbf{23^{\circ}C}$, $\mathbf{30^{\circ}C}$ and $\mathbf{37^{\circ}C}$.}

\begin{figure}[H]
  \centering
      \makebox[\textwidth]{\includegraphics[width=0.95\textwidth]{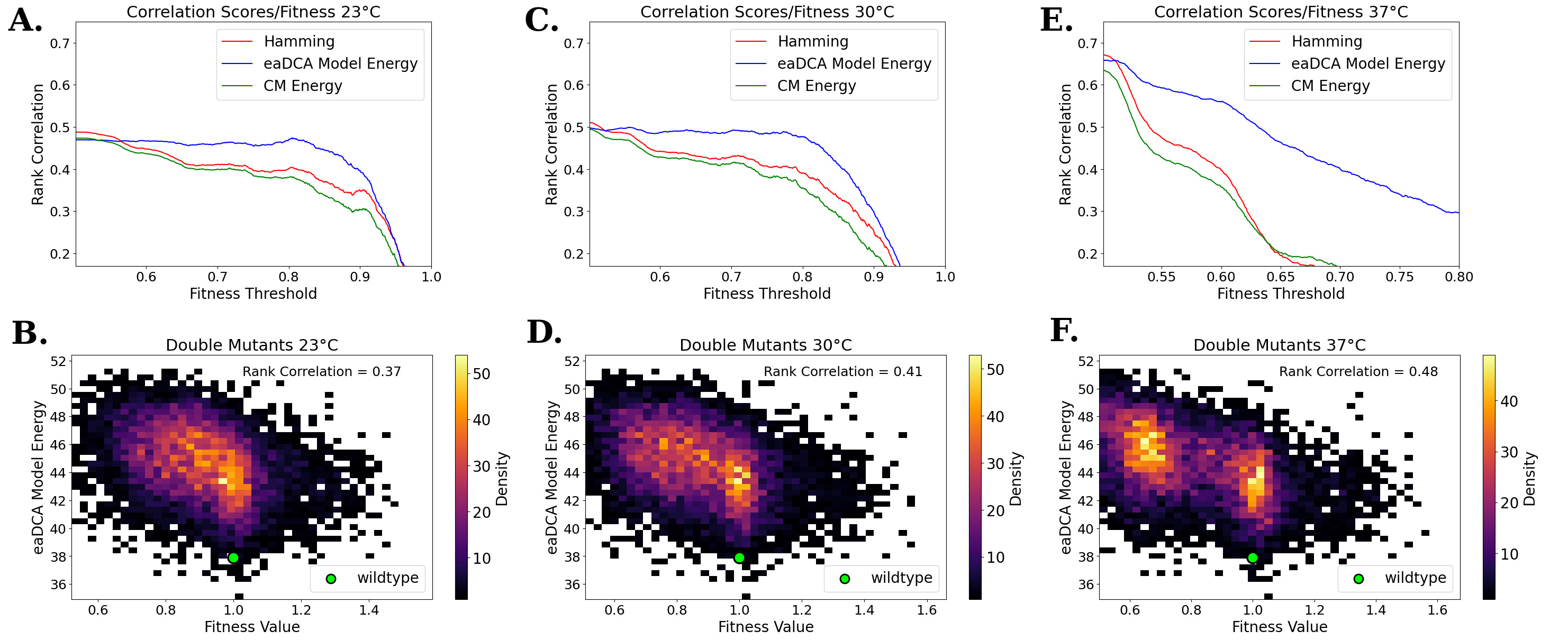}} 
  \caption{\textbf{A.} Correlation of Hamming distance, eaDCA model energy and CM
energy with tRNA  $23^{\circ}C$ fitness at different values of minimum fitness threshold. \textbf{B.} Relation between eaDCA model energy and tRNA $23^{\circ}C$ fitness on the $8101$ double
mutants. \textbf{C.} Correlation of Hamming distance, eaDCA model energy and CM
energy with tRNA  $30^{\circ}C$ fitness at different values of minimum fitness threshold. \textbf{D.} Relation between eaDCA model energy and tRNA $30^{\circ}C$ fitness on the $8101$ double
mutants. \textbf{E.} Correlation of Hamming distance, eaDCA model energy and CM
energy with tRNA  $37^{\circ}C$ fitness at different values of minimum fitness threshold (also present in the main text). \textbf{F.} Relation between eaDCA model energy and tRNA $37^{\circ}C$ fitness on the $8101$ double
mutants (also present in the main text).}
  \label{fig:example}
\end{figure}

\section{SHAPE-MaP Experiments}

\subsection{Selection Process For The Tested tRNA}

We delineate the two-step selection process employed to select the $76$ tested tRNA sequences generated from the eaDCA model. \\
To begin, $12000$ artificial tRNA sequences were generated based on the RF00005 model. During this generation process, we kept the last $16$ residues fixed to the yeast tRNA(asp) ones, while also prohibiting the introduction of gap states. \\
The selection process involved two filtering criteria:\\

\noindent \textbf{-} Criteria 1: Secondary Structure 

\begin{itemize}
\item The secondary structure of yeast tRNA(asp) was predicted using the RNAfold tool from the Vienna RNA (October 2022) package.
\item  F score $F$ was calculated between the predicted secondary structure and the consensus tRNA secondary structure.
\item The pool of $12000$ sequences was reduced, retaining only the sequences whose RNAfold-predicted structure had an f-score greater than that of yeast tRNA(asp) ($F >53)$ .
\end{itemize}

\noindent \textbf{-} Criteria 2: eaDCA energy \\

The secondary structure-filtered dataset was further refined based on the energy of the eaDCA model:

\begin{itemize}
\item The $38$ sequences of group A were randomly selected among the ones that had an energy lower than that of yeast tRNA(asp) ($E(a_1,\dots,a_L) < 44$).
\item The $38$ sequences of group B were randomly selected among the ones that had an energy lower than the $6\%$ left tail of the energy distribution ($E(a_1,\dots,a_L) < 35$).
\end{itemize}

This two-step filtering process based on secondary structure and model energy, therefore, resulted in two distinct groups, Group A and Group B, each composed of 38 tRNA sequences.

\begin{figure}[H]
  \centering
      \makebox[\textwidth]{\includegraphics[width=0.9\textwidth]{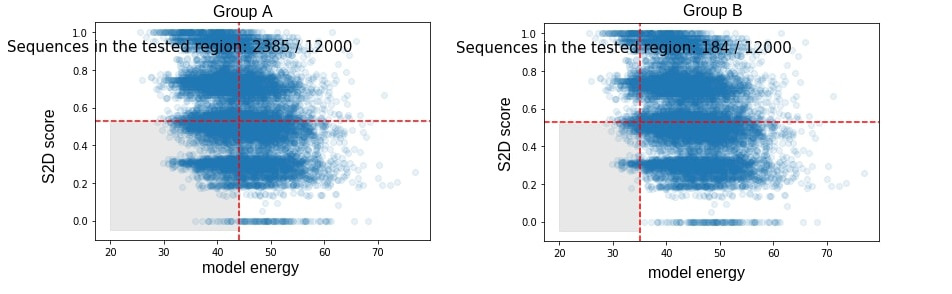}} 
  \caption{To visualize the filtering criterion applied, we shaded in graythe areas from which samples for Group A and Group B were randomly selected.}
  \label{fig:example}
\end{figure}

 \newpage
 
\subsection{Tested tRNA Dataset Characteristics}

\subsubsection{Sequence Diversity}
\begin{figure}[H]
  \centering
      \makebox[\textwidth]{\includegraphics[width=0.9\textwidth]{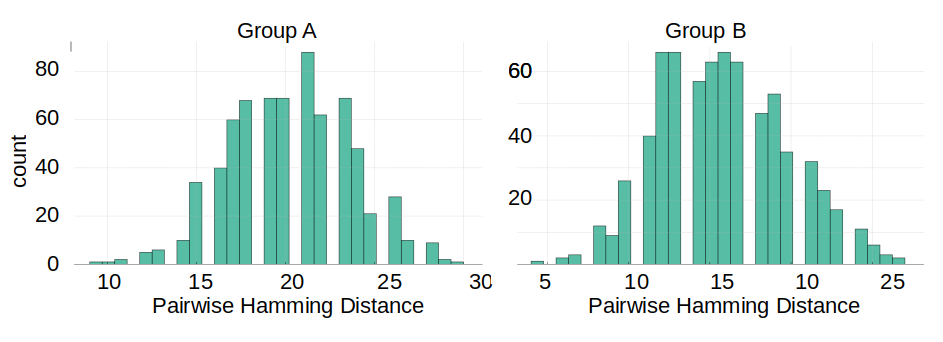}} 
  \caption{Intra-group pairwise sequence distances. This represents the \textit{diversity} of the datasets. Group B, as expected, is less diverse since the energy filtering criterion is more stringent.}
  \label{fig:example}
\end{figure}

\subsubsection{Sequence Novelty}

\begin{figure}[H]
  \centering
      \makebox[\textwidth]{\includegraphics[width=1\textwidth]{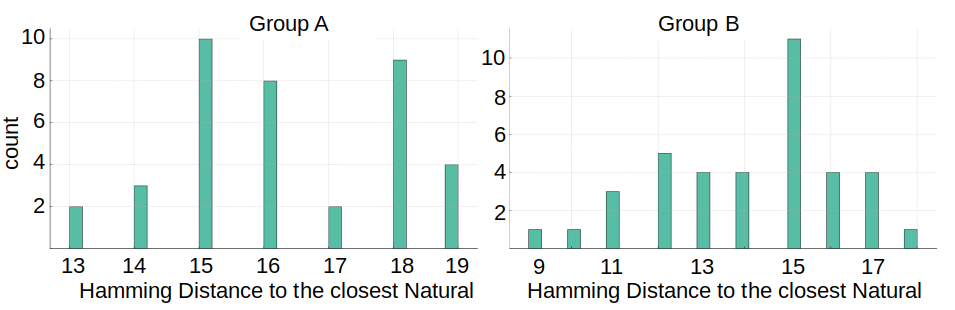}} 
  \label{fig:example}
   \caption{Distance from the closest natural for the two groups. This represents the \textit{novelty} of the datasets. Group B, as expected, introduces less novelty since the energy filtering criterion is more stringent. It's important to note that in our generation, the last 16 sites are maintained as constant, matching those of the yeast tRNA-Asp. These sites are not only conserved but also participate in forming secondary structure pairs. Consequently, effectively 32 out of 71 sites are fixed in our investigations. The novel aspects of our study are primarily concentrated on the remaining 39 variable sites.}
\end{figure}

\newpage 

\subsection{SHAPE-MaP Detailed Protocol}

\subsubsection{RNA Production}
The constructs were designed with the T7 promoter positioned at the 5' end of the tRNAs, and with the 15 terminal nucleotides of each tRNA intentionally fixed to facilitate PCR amplification. In total, 77 tRNAs were designed. Out of these, 7 tRNAs were individually produced and modified as single RNA using gBlock DNA templates from Integrated DNA Technologies. The remaining tRNAs were produced and modified as pools of 10 tRNAs using oligoPools from Integrated DNA Technologies.

The DNA templates were amplified by PCR, using 0.5 uM of forward and reverse primers, 0.2 mM of dNTPs (Sigma), 1U of Phusion Hot Start Flex polymerase (New England Biolabs), 1 ng of DNA template, and 1X HF Phusion buffer. The resulting amplified DNA was purified using the NucleoSpin Gel and PCR cleanup kit from Macherey Nagel and quantified using a Nanodrop spectrophotometer. The in vitro transcription was carried in 20 ul reaction volume for 4 h at 37 C, using the HiScribe T7 High Yield RNA Synthesis Kit (New England Biolabs) according to manufacturer instructions, with 800 to 1200 ng of DNA.The transcribed RNAs were then purified by phenol-chloroform extraction using cold acid phenol-chloroform at pH 4.5 (Invitrogen) and ethanol precipitation with 0.1 volume of 3M Sodium Acetate (Sigma) and 2.5 volumes of ice cold 100\% ethanol. The precipitated RNAs were resuspended in 50 ul of nuclease-free water and treated with 5U of DNase I (New England Biolabs) for 15 min at 37°C. The RNAs were mixed with loading buffer containing 70\% formamide, 130 mM EDTA, 0.1\% xylene cyanol, and 0.1\% bromophenol blue and loaded onto a pre-run 8\% urea polyacrylamide gel electrophoresis (PAGE). The RNAs were then purified from the gel, ethanol precipitated, resuspended in nuclease-free water, quantified using the Qubit RNA Broad Range assay kit (ThermoFisher) and diluted to a concentration of 10 uM.

\subsubsection{RNA Modification}

5 pmol of RNA in 6 ul of RNase-free water were heated at 95°C for 2 min and placed on ice for 3 min. Subsequently, 3 ul of 3X folding buffer (final concentration: 50 mM Hepes pH 8.0, 200 mM potassium acetate pH 8.0, and 3 mM MgCl2) were added, and the tubes were incubated at 37°C for 20 min to allow the refolding of RNAs. 1 ul of 10X 1M7 (Sigma) in DMSO for the positive reaction (final concentration : 10 mM 1M7) or 1 ul of neat DMSO for the negative reaction was added and rapidly mixed to ensure even distribution. The solutions were immediately incubated at 37°C on a heating block and the reaction was allowed to proceed for 5 min. Following the modification step, the RNA solutions were cooled on ice and purified by ethanol precipitation. For the denaturing condition, 5 pmol of RNA in 3 ul in RNase-free water were mixed with 5 ul of 100\% highly deionized formamide (Hi-Di Formamide, Applied Biosystems) and 1 ul of 10X denaturation buffer (final concentration : 50 mM Hepes pH 8.0, 40 mM EDTA). The tubes were incubated at 95°C for 1 min to denature the RNAs. Similarly, 1 ul of 10X SHAPE reagent in DMSO was added to the denatured RNAs, rapidly mixed, and immediately incubated at 37°C for 5 min. After the modification step, the solutions were cooled on ice and purified alongside the negative and positive reactions.

The ethanol precipitation was performed as follows: the 10 ul of modified RNAs were diluted with 87 ul of RNase-free water, and mixed with 1 ul of 20 mg/ml glycogen (Invitrogen) and 2 ul of 100 mM EDTA pH 8. Next, 0.1 volume of 3 M NaAc and 3.5 volumes of ice cold 100\% ethanol were added to the solution, and the RNas were precipitated overnight at -20°C, centrifuged at 13,000 rpm for 1 h at 4°C, and resuspended in 15 ul of RNase-free water. The RNA concentrations were quantified using the Qubit RNA High Sensitivity assay kit (ThermoFisher).

\subsubsection{Library Preparation}

After purification, the modified RNAs were pooled in equimolar proportions based on their conditions (positive, negative, and denaturing) for reverse transcription. The reverse-transcription was carried out with 1 pmol of RNA in 20 ul of reaction volume, as follows : i) the RNAs were incubated at 95°C for 3 min and transferred to ice, ii)  1 ul of a 2 uM reverse primer, complementary to the end of the tRNA and carrying the Rd2 adaptor, was added to the solutions and incubated at 65°C for 5 minutes and transferred to ice, iii) 8 ul of freshly prepared 2.5X Shape-Map buffer (final concentration: 50 mM Tris-HCl pH 8.3, 75 mM KCl, 10 mM DTT, 6 mM MnCl2, 0.5 mM dNTP mix) was added and incubated for 2 min at 50°C before being placed on ice, iv)  200 U of SuperScript II reverse transcriptase (ThermoFisher) and 1 ul of 100 uM Rd1-TSO were added and incubated at 42°C for 3 hours, followed by an inactivation step at 70°C for 15 minutes. The resulting cDNAs were purified with AMPure XP beads (Beckman Coulter) with a bead-to-sample ratio of 1.8, and eluted in 15 ul of RNase-free water.

A PCR enrichment was carried out for 8 cycles, using 25 ul of KAPA HiFi HotStart ReadyMix (Roche), 5 ul of each primer (NEBNext Multiplex Oligos for Illumina, New England Biolabs), and 15 ul of purified cDNAs. 
The DNA libraries were purified with AMPure XP beads (Beckman Coulter) with a bead-to-sample ratio of 0.9, and eluted in 20 ul of RNase-free water, and quantified using the Qubit dsDNA High Sensitivity assay kit (ThermoFisher) and quantitative PCR (KAPA Library Quantification Kit, Roche). 
The length distribution of the libraries was analyzed using the High Sensitivity D1000 ScreenTape for TapeStation Systems (Agilent). 
The libraries were sequenced on a MiSeq-V3 flow cell (Illumina) in paired ends with 25\% of PhiX by the NGS platform at Institut Curie (Paris, France).

\begin{table}[h]
\centering
\begin{tabular}{l p{0.75\textwidth}}
\hline
\textbf{Primer Name} & \textbf{Sequence (5'-3')} \\
\hline
tRNA\_R & TGCCGCGACGGGGAA \\
tRNA\_F & TAATACGACTCACTATAG \\
TSO\_Rd1 & TACACGACGCTCTTCCGATCTrGrGrG \\
Rd2\_tRNA & GTGACTGGAGTTCAGACGTGTGCTCTTCCGA-TCTTGCCGCGACGGGGAA \\
i5 primer & AATGATACGGCGACCACCGAGATCTACACTC-TTTCCCTACACGACGCTCTTCCGATC-s-T \\
i7 primer & CAAGCAGAAGACGGCATACGAGAT-NNNNNN-GTGACTGGAGTTCAGACGTGTGCTCTTCCGATC-s-T \\
\hline
\end{tabular}
\caption{Sequences of the primers used. -s- indicates a phosphorothioate bond. N indicates random nucleotide}
\end{table}

\clearpage

\subsubsection{SHAPE Reactivity Profiles}

Here we show the projection of SHAPE reactivity onto the tRNA consensus secondary structure for tested sequences (and for another natural tRNA took from the SHAPE reference dataset).  There doesn't appear to be a significant difference between groups A and B among the sequences that exhibit a compatible profile.

\clearpage 

\begin{figure}[H]
  \centering
  \makebox[\textwidth]{\includegraphics[height=1\textheight]{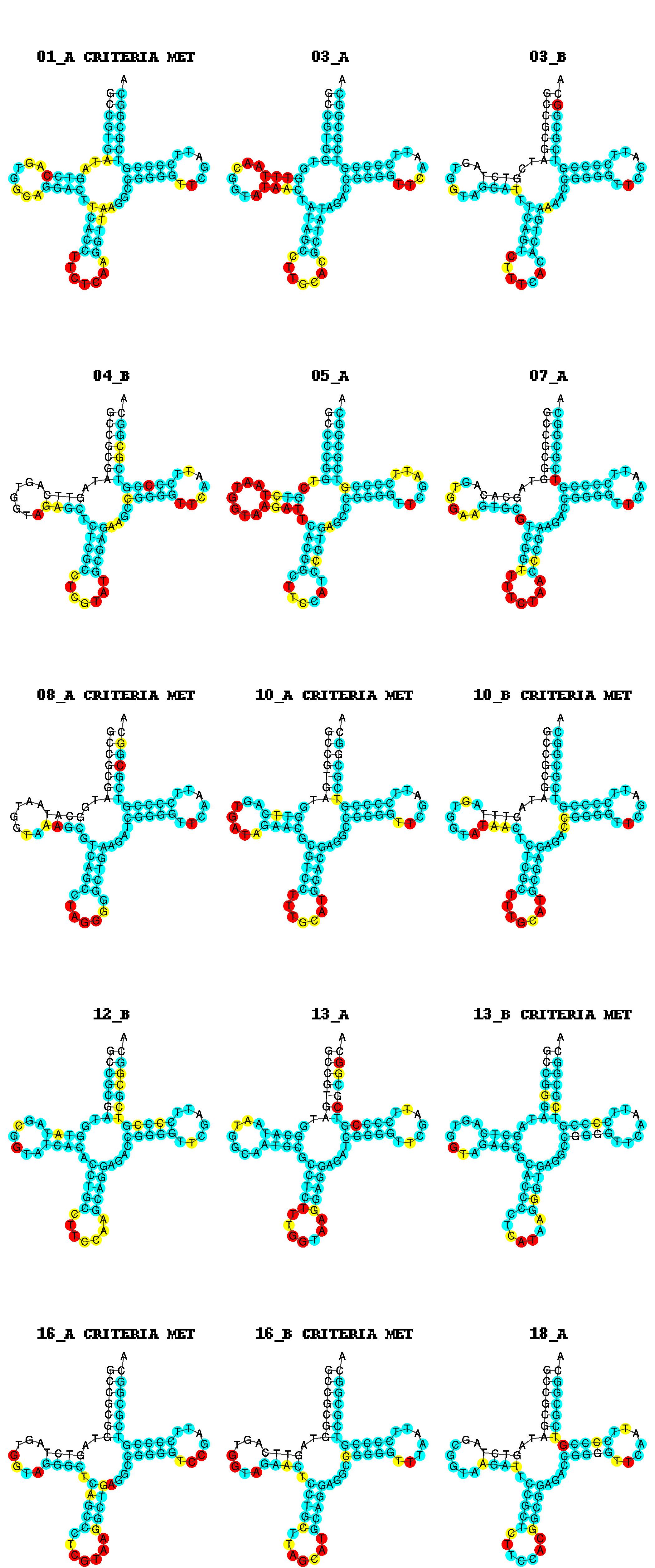}} 
\end{figure}

\begin{figure}[H]
  \centering
  \makebox[\textwidth]{\includegraphics[height=1\textheight]{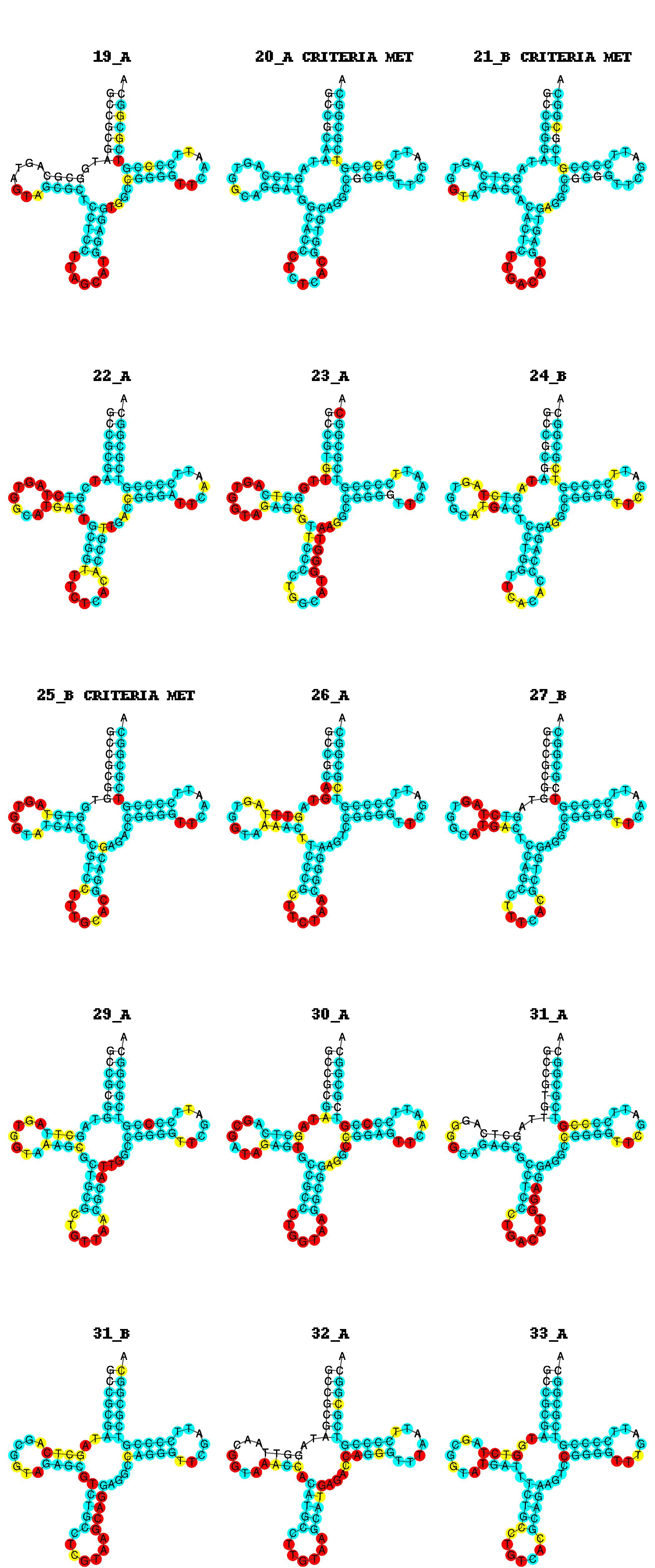}} 
\end{figure}

\begin{figure}[H]
  \centering
  \makebox[\textwidth]{\includegraphics[height=1\textheight]{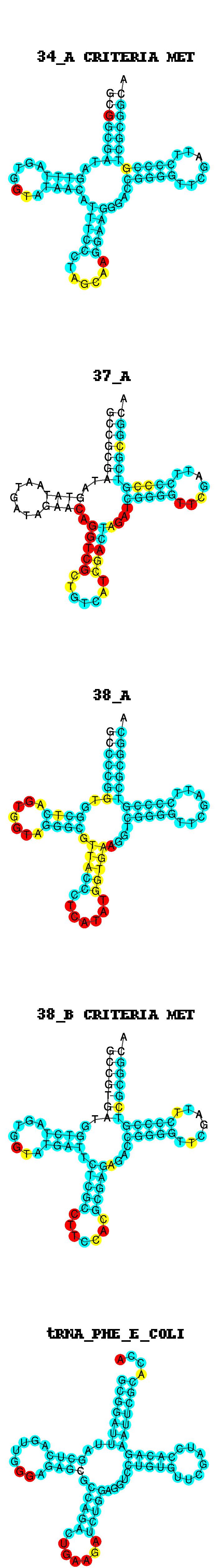}} 
\end{figure}

\end{document}